\documentclass{mn2e}
\usepackage{graphicx}

\newif\ifAMStwofonts

\input epsf

\def\lesssim{\mathrel{\hbox{\rlap{\hbox{\lower4pt\hbox{$\sim$}}}\hbox{$<$}}}}

\def\gtrsim{\mathrel{\hbox{\rlap{\hbox{\lower4pt\hbox{$\sim$}}}\hbox{$>$}}}}

\def\teff{$T_{\rm eff}$~}

\def\ll_lsun{$\log({L/\rm L_{\odot}})$~}

\def\masa_msun{$M/ \rm M_{\odot}$~}

\def\m_mstar{$M/M_{*}$~}

\title{Evolution and colours of helium-core white dwarf stars: the case of low 
metallicity progenitors}

\author [A.  M.  Serenelli, L.  G.  Althaus, R.  D.  Rohrmann
and O. G. Benvenuto] 
{A.  M. Serenelli$^1$
\thanks{Fellow of the Consejo Nacional de Investigaciones
Cient\'{\i}ficas   y    T\'ecnicas   (CONICET),   Argentina.    Email:
serenell@fcaglp.unlp.edu.ar},  
L.  G. Althaus$^1$
\thanks{Member  of the Carrera  del  Investigador  Cient\'{\i}fico y  
Tecnol\'ogico,  Consejo Nacional de  Investigaciones Cient\'{\i}ficas y  
T\'ecnicas (CONICET), Argentina.      Email:     althaus@fcaglp.unlp.edu.ar},    
R. D. Rohrmann$^{2,3}$
\thanks{Postdoctoral Fellow of the IA-UNAM.    Email:
rohr@astroscu.unam.mx} and 
O. G. Benvenuto$^{1}$\thanks{Member  of the Carrera  del  Investigador  
Cient\'{\i}fico, Comisi\'on de Investigaciones Cient\'{\i}ficas de la 
Provincia de Buenos Aires. Email: obenvenu@fcaglp.unlp.edu.ar} 
\\  $^1$Facultad  de  Ciencias  Astron\'omicas  y
Geof\'{\i}sicas de la Universidad Nacional de  La Plata and 
Instituto Astrof\'{\i}sica La Plata (CONICET), \\ Paseo  del Bosque
S/N,  (1900) La  Plata, Argentina  \\  $^2$Observatorio Astron\'omico,
Universidad  Nacional  de C\'ordoba,  Laprida  854, (5000)  C\'ordoba,
Argentina\\ $^3$Present address:  Instituto de Astronom\'{\i}a, UNAM, 
A.P. 70-264, 04510 M\'exico D.F., M\'exico}

\pagerange{\pageref{firstpage}--\pageref{lastpage}}

\pubyear{2002}

\begin{document}

\maketitle

\label{firstpage}

\begin{abstract} 

The present work  is designed to explore the  evolution of helium-core
white dwarf  (He WD)  stars for the  case of metallicities  much lower
than the solar metallicity ($Z$= 0.001 and $Z$= 0.0002).  Evolution is
followed in a self-consistent way with the predictions of detailed and
new non-grey  model atmospheres, time-dependent  element diffusion and
the history  of the white  dwarf progenitor.  Reliable  initial models
for low mass  He WDs are obtained  by applying mass loss rates  to a 1
M$_\odot$ stellar model in such  a way that the stellar radius remains
close to the Roche lobe radius. The loss of angular momentun caused by
gravitational  wave emission  and  magnetic stellar  wind braking  are
considered.  Model  atmospheres, based on  a detailed treatment  of the
microphysics  entering the  WD atmosphere  (such as  the  formalism of
Hummer-Mihalas  to deal with  non-ideal effects)  as well  as hydrogen
line  and pseudo-continuum  opacities,  enable us  to provide  accurate
colours and magnitudes at both early and advanced evolutionary stages.

We  find that most  of our  evolutionary sequences  experience several
episodes of  hydrogen thermonuclear flashes. In  particular, the lower
the metallicity, the larger the minimum stellar mass for the occurrence
of  flashes  induced  by  CNO  cycle  reactions.  The
existence of a mass  threshold for the occurrence of diffusion-induced
CNO flashes  leads to  a marked  dichotomy in the  age of  our models.
Another  finding of this  study is  that our  He WD  models experience
unstable  hydrogen burning via  PP nuclear  reactions at  late cooling
stages as a  result of hydrogen chemically diffusing  inwards. Such PP
flashes take place in models with very low metal content.

We also find  that  models
experiencing CNO flashes exhibit  a pronounced turn-off in  most of
their colours at M$_{\rm V} \approx$ 16.Finally, colour-magnitude  diagrams for  our models are  presented and
compared with  recent observational  data of He  WD candidates  in the
globular  clusters  NGC 6397  and  47  Tucanae.  

\end{abstract}


\section{Introduction} \label{sec:intro}

Low-mass helium-core white dwarf stars  (He WDs) are thought to be the
result of the evolution of certain close binary systems. Mass-transfer
episodes  in binary  systems  are required  to  form low  mass He  WDs
because an isolated  star would need a timescale  much longer than the
present  age  of the  universe  to reach  a  WD  configuration with  a
helium-rich interior. Low mass WDs have been detected in large surveys
(Bragaglia et  al. 1990; Bergeron, Saffer \&  Liebert 1992; Bragaglia,
Renzini  \&  Bergeron  1995;  Saffer,  Livio \&  Yungelson  1998)  and
represent an appreciable fraction of the total population of WD stars.

Since  He WDs  began to  be  found in  numerous binary  configurations
(Marsh 1995; Marsh, Dhillon \& Duck 1995; Lundgren et al. 1996; Moran,
Marsh  \& Bragaglia  1997; Orosz  et al.   1999; van  Kerkwijk  et al.
2000), they have  captured the attention of many  researchers who have
devoted much effort to their study.  Recent works with the emphasis on
the  evolutionary  properties  of  these stars  include  Benvenuto  \&
Althaus (1998), Hansen \& Phinney (1998), Driebe et al. (1998), Sarna,
Ergma \& Antipova (2000),  Althaus, Serenelli \& Benvenuto (2001a) and
Serenelli et al.  (2001).  In particular, Althaus et al.  (2001a) have
explored  their  evolution  in  a  self-consistent  way  with  nuclear
burning, time  dependent element diffusion  and the history of  the WD
progenitor.   Althaus  et al.   (2001a)  find  that element  diffusion
induces thermonuclear  hydrogen shell flashes  in He WDs  with stellar
masses greater  than $\approx$  0.18 M$_\odot$.  As  a result,  He WDs
more massive  than 0.18 M$_\odot$  are characterized by  thin hydrogen
envelopes and a fast evolution,  while less massive He WDs evolve much
more  slowly. As  shown by  Althaus  et al.   (2001a), this  behaviour
solves the discrepancies between the spin-down ages of the millisecond
pulsars B1855+09,  PSR J0034-0534 and  PSR J1012+5307 and  the cooling
ages of their  He WD companions.  The evolution of  the Althaus et al.
(2001a)   stellar   models   in   the   colour-colour   and   in   the
colour-magnitude  diagrams  have been  analysed  by  Serenelli et  al.
(2001) who find, on the  basis of detailed non-grey model atmospheres,
that the  emergent spectrum  of low mass  He WDs becomes  bluer within
time-scales of  astrophysical interest when  the effective temperature
decreases  below  4000K.   Because   Serenelli  et  al.   (2001)  were
interested in the late stages of He WD evolution, they did not attempt
a detailed  modeling of the emergent  spectrum of these  stars at high
effective temperature stages.

Interestingly, He WDs have also  been detected or inferred in open and
globular clusters (Anderson 1997; Landsman et al. 1997; Edmonds et al.
1999).  More recently, Edmonds  et al.  (2001) have optically detected
the  He  WD companion  to  a millisecond  pulsar  in  47 Tucanae.   In
addition, Taylor et al. (2001)  have presented evidence for a sequence
of  He WD  canditates  in the  globular  cluster NGC  6397.  A  proper
interpretation  of the  observations of  He WDs  in  globular clusters
requires evolutionary  calculations for HeWD progenitors with much 
lower metallicities than the solar one. 
In  this  regard,   the  present  work  is  designed   to  extend  the
evolutionary calculations presented in Serenelli et al.  (2001) to the
case  of low  metallicities.   In addition,  the present  calculations
constitute  an improvement over  those presented  in Serenelli  et al.
(2001).  Here  a more detailed treatment of  the microphysics entering
the  WD  model  atmosphere  than   that  presented  in  that  work  is
considered, thus enabling us to derive accurate colours and magnitudes
for He  WDs at high effective  temperatures where the  effects of line
broadening  opacities  are  not  negligible.   Finally,  a  much  more
realistic treatment of mass loss phases than that attempted in Althaus
et al. (2001a) and Serenelli et al. (2001) is considered in the present
study. Details  about our atmosphere models, evolutionary code 
and mass  loss treatment are briefly  described in  Section 2.   Results are
presented in Section  3 and  Section 4 is devoted
to making some concluding remarks.

\section{Computational details and input physics}

In this  section we  describe the main  characteristics and  the input
physics  of  both  the   model  atmosphere  and  evolutionary  codes.
Computational details  concerning calculation of  pre-WD evolution are
also presented.

\subsection{Model atmosphere code}

At  low  effective  temperatures,  WD  cooling  is  sensitive  to  the
treatment  of the  outer boundary  condition. Thus,  appropriate outer
boundary conditions  for the cooling He  WD models are  derived on the
basis of detailed non-grey model  atmospheres, as done in Serenelli et
al. (2001).   This is an important aspect  as far as cool  WD ages are
concerned. Additionally, these model atmospheres allow us to provide a
grid of  colour indices  and magnitudes for  our evolving  models. The
model  atmospheric  structure  is  described  at  length  in  Rohrmann
(2001). Briefly,  it is based  on the assumption of  constant gravity,
local  thermodynamic  equilibrium  and  plane-parallel  geometry,  and
includes the  hydrogen and  helium species (zero  metallicity). Energy
transfer  by radiation  and convection  is considered  and  a standard
linearization technique is employed  to solve the resulting equations.
The following species are taken  into account: H, H$_2$, e$^-$, H$^-$,
H$^+$,  H$_2$$^+$,  H$_3$$^+$, He,  He$^+$  and He$^{++}$.    
All  relevant  bound-free,  free-free  and  scattering
processes  contributing   to  opacity   have  been  included   in  our
calculations.  Up-to-date collision-induced absorptions (CIA opacities) 
due  to H$_2$-H$_2$, H$_2$-He
and H-He collisions  are taken from the cross  section calculations by
Borysow,  Jorgensen  \&  Fu  (2001),  Jorgensen  et  al.   (2000)  and
Gustafsson \&  Frommhold (2001), respectively. CIA  represents a major
source  of opacity  in the  infrared and  dominates the  shape  of the
emergent  spectrum at  low effective  temperatures.   Broadband colour
indices have  been calculated using  the optical $UBVRI$  and infrared
$JHKL$  passbands  of Bessell  (1990)  and  Bessell  \& Brett  (1988),
respectively.   We use the  callibration constants  as derived  on the
basis of the synthetic flux of  the model (\teff =9400 K, $\log g =
3.95$) calculated by Kurucz (1979) for Vega, rather than the constants
of Bergeron, Ruiz \& Leggett (1997) we employed in Serenelli et al. (2001).

In the present work,  important improvements have been incorporated in
our model atmospheres, which have enabled us to derive accurate colours
and  magnitudes for  both early  and advanced  stages of  the cooling.
Specifically,  hydrogen  line opacities  from  the  Lyman, Balmer  and
Paschen series and pseudo-continuun  absorptions have been included in
our  calculations.   Also, non-ideal  effects  in  the computation  of
chemical equilibrium of mixed  hydrogen and helium gases are accounted
via  the occupation formalism  of Hummer  \& Mihalas  (1988), together
with  a modified  version of  the  optical simulation  of D\"appen, 
Anderson \& Mihalas (1987) to  assess opacities in non-ideal gases.   
For more details
about the model  atmospheres used in this work we  refer the reader to
Rohrmann et al. (2002).

\begin{figure}
\includegraphics[width=8.5cm]{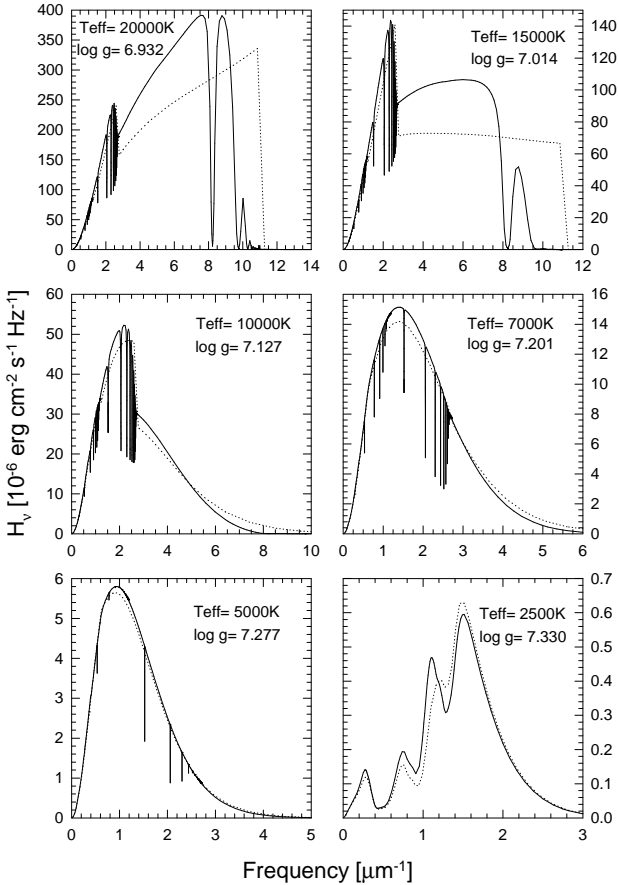}
\vskip -.5cm
\caption{Emergent spectra  calculated for some selected  models of our
evolutionary  sequence of  0.280~M$_\odot$ and  $Z$=  0.0002.  Effective
temperatures and surface gravities are indicated in each panel. Models
are  characterized  by  pure  hydrogen outer  layers,  resulting  from
element  diffusion.    Solid  lines  show   the  present  calculations
including  line  opacity,  non-ideal  gas  model  and  up-to-date  CIA
opacities.   Dotted  lines  indicate  the results  obtained  with  our
previous treatment of model atmospheres.}
\label{fig.flu} 
\end{figure}

In   Fig.~\ref{fig.flu}  we   show  the   emergent  spectrum   of  our
0.280~M$_\odot$  He   WD  model  at   some  selected  points   of  its
evolution. The  prediction of  the present calculations,  plotted with
solid lines, are compared with  our older treatment given in Serenelli
et al.  (2001).  Note  that for effective temperatures above $\approx$
7000K,  hydrogen   lines  strongly  modify  the  shape   of  the  flux
distribution, so their effect is expected to affect particularly the U
and B colours. At low  effective temperature, the emergent spectrum is
modeled by  the CIA opacity,  and the bulk  of radiation is  forced to
emerge  at   higher  frequencies   as  compared  with   the  blackbody
prediction. As a result, most colour indices will exhibit a pronounced
turnoff  to the  blue  as the  effective  temperature decreases  below
$\approx$ 4000K. Note finally  that the present calculation predicts a
flux  distribution   somewhat  different   from  that  of   our  older
treatment. This is due partly to the inclusion of nonideal effects and
the use of updated CIA opacity data as well.

\subsection{Evolutionary code and pre-white dwarf evolution}

The evolutionary calculations presented in this work have been carried
out  with the  same evolutionary  code  that described  in Althaus  et
al. (2001a).  We present briefly  here the main physical inputs of the
code and  refer the  reader to that  article for a  more comprehensive
description.  WD  evolution is  accomplished in a  self-consistent way
with abundance  changes as predicted by element  diffusion and nuclear
burning.   In  order  to  achieve this,  gravitational  settling,  and
chemical  and thermal  diffusion of  nuclear species  have  been fully
taken  into  account following  the  multicomponent  gas treatment  of
Burgers  (1969). Hydrogen  burning  is considered  through a  complete
network of thermonuclear reactions for the proton-proton chain and the
CNO bi-cycle.   As a result  of the various  diffusion processes, the metallicity 
distribution evolves with time; thus  radiative  opacities  are  
calculated  consistently  with
predictions  from element diffusion  and are  taken from  OPAL opacity
tables (Iglesias \& Rogers 1996).

Initial He  WD models  have been obtained  by abstracting mass  from a
1~M$_\odot$ model at appropriate stages  of its evolution off the main
sequence. We have considered  two different metallicity values for our
WD progenitor  models, $Z$= 0.001 and  $Z$= 0.0002, chosen in  order to be
representative of low metallicity  environments that are found in many
globular clusters.   For He WD models more  massive than 0.2~M$_\odot$
in the  case of  a metallicity  value $Z$= 0.001  and more  massive than
0.23~M$_\odot$ for $Z$=0.0002, mass loss was invoked at different stages
during the  evolution along the red  giant branch. 
However,  for  less  massive  models  we  have  followed  a  different
procedure.  Mass  loss is initiated  soon after the  1~M$_\odot$ model
leaves  the main  sequence and  starts its  excursion redwards  in the
Hertzsprung-Russell diagram (HRD). We  have followed the accepted idea
that  He WDs  (particularly  low mass  He  WDs) are  formed in  binary
systems where the He WD  progenitor undergoes a strong mass loss phase
due  to Roche  lobe  overflow. Indeed,  for  low mass  models we  have
computed pre-WD evolution  by imposing that mass loss  rates are fixed
by  the condition that  the stellar  radius must  remain close  to the
radius of the Roche lobe.  To  compute the radius of the Roche lobe we
assume  that  our  1~M$_\odot$  model has  a  1.4~M$_\odot$  companion
(representative of a pulsar  companion). Given the initial semiaxis of
the orbit,  we obtain  the radius of  the Roche  lobe by means  of the
analytic approximation given by Eggleton (1983)
\begin{equation}
r_L = a \frac{0.49 q^{2/3}}{0.6 q^{2/3} + ln(1+q^{1/3})},
\label{eq.egg}
\end{equation}
where $r_L$ is  the radius of the Roche lobe, $a$  the semiaxis of the
orbit and $q$  is the mass ratio. We have assumed  that mass loss from
our 1~M$_\odot$  is fully non-conservative,  i.e. all the  matter lost
from the WD progenitor leaves the system and carries off its intrinsic
angular momentum. To this end,  we have followed the formulation given
by Podsiadlowski, Joss \& Hsu (1992), that is
\begin{equation}
\frac{\dot J_{ML}}{J} = \frac{M}{M_1 M_2} {\dot M_1} {\rm yr^{-1}}.
\end{equation}
Here, $J$  means the  total angular momentum,  $\dot J_{ML}$  its time
derivative  caused by mass  loss and $M$  the total  mass of  the system.
Sub-indices 1  and 2 refer to  the WD progenitor  and the 1.4$M_\odot$
companion respectively.  Finally  $\dot M_1$ is the mass  loss rate of
the  WD progenitor.   In  addition, we  have  also considered  angular
momentum  losses  due  to  gravitational  wave  radiation  (Landau  \&
Lifshitz 1971)
\begin{equation}
\frac{\dot J_{GR}}{J}  = - 8.5  \times 10^{-10} \frac{M_1  M_2 M}{a^4}
{\rm yr^{-1}},
\end{equation}
which  is an  important  contribution to  angular  momentum losses  in
binaries  with very  short  periods.  Finally,  magnetic stellar  wind
braking when  a convective envelope developes has  also been accounted
for as in Sarna et al. (2000):
\begin{equation}
\frac{\dot J_{MB}}{J}  = - 3  \times 10^{-7} \frac{M^2  R^2_1}{M_1 M_2
a^5} {\rm yr^{-1}}.
\end{equation}
In  order  to follow  the  time evolution  of  $J$,  we integrate  the
equation
\begin{equation}
\frac{\dot  J}{J}=  \frac{\dot J_{ML}}{J}  +  \frac{\dot J_{GR}}{J}  +
\frac{\dot J_{MB}}{J}
\end{equation}
and with the  aid of Kepler's third law we finally  get $a$ (and $r_L$
through Eq.~\ref{eq.egg}).  The present treatment for mass loss is not
self-consistent since the mass loss  rate is not incorporated as a new
unknown quantity  to be solved  during the iteration procedure  in the
Henyey  scheme used  to  solve the  stellar  structure equations,  but
instead is fixed in advance for each model. However, we want to recall
that our main  interest in this work is not  to analyse the properties
(initial/final periods, masses, semiaxis)  of binary systems that lead
to the  formation of HeWD-millisecond  pulsar pairs, but to  study the
cooling  time-scales  and  photometric  properties of  cooling  HeWDs.
Also, it is worth mentioning that the algorithm applied in the present
work to simulate mass loss phases represents a much better approach to
obtain  physically  sound initial  HeWD  models  than  that of  simply
abstracting mass as done previously in Althaus et al. (2001a).

\begin{figure}
\includegraphics[width=8.25cm]{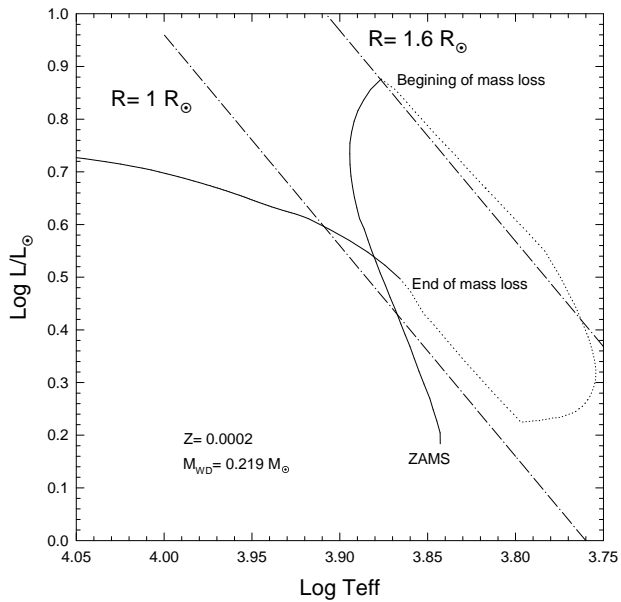}
\vskip -3.5cm
\caption{Hertzsprung-Russell   diagram  for  the   0.219~M$_\odot$  WD
progenitor  with  initially  1~M$_\odot$  and $Z$=0.0002.   Dotted  line
depicts the  evolution during mass  loss phase, which starts  when the
star fills its Roche lobe  at R= 1.6~R$_\odot$.  The initial and final
semiaxis  are 5.25  and 4.6~R$_\odot$  respectively.  The  initial and
final periods are 0.9d in both cases.}
\label{fig.bin}
\end{figure}

For the  purpose of illustration,  we show in Fig.   \ref{fig.bin} the
evolution  in the  HRD for  the 0.219~M$_\odot$  WD  progenitor having
initially  1~M$_\odot$.  Note  that during  the first  stages  of mass
loss,  the WD  progenitor evolves  at almost  constant  radius.  Small
deviations from constant-radius evolution  are caused by the fact that
the  semiaxis increases slightly  as a  result of  mass loss  from the
least massive component  of the pair. As evolution  proceeds, an outer
convection  zone developes  and  magnetic braking  starts  to play  an
important role  with the  consequent result that  the orbit  starts to
shrink. During the final stage of mass loss, the WD progenitor evolves
again at constant radius until  it eventually shrinks within the Roche
lobe and mass loss halts. To  place our results on a more quantitative
basis we list in Table~\ref{tab.bin} the main characteristics for some
of our  less massive  WD progenitors.  In  particular, for a  given WD
model, we list the metallicity,  the surface hydrogen abundance at the
end of mass loss phase and  the initial and final period and semiaxis.
We  want  to mention  that  our results  agree  quite  well with  more
detailed  calculations in  which  the mass  loss  phase is  calculated
self-consistently (Sarna et al. 2000).

\begin{table}
\begin{center}
\caption{Main  characteristics  of  some  low mass  pre-WD  progenitor
calculations.}
\begin{tabular}{@{}c@{\extracolsep{3mm}}r@{\extracolsep{0mm}.}l@{\extracolsep{3mm}}
c@{\extracolsep{3mm}}c@{\extracolsep{3mm}}c
@{\extracolsep{3mm}}c@{\extracolsep{3mm}}c}    \hline    M$_{\rm   f}$
[M$_\odot$] & \multicolumn{2}{c}{Z} & X$_{\rm  s}$ & P$_{\rm i}$ [d] &
$a_{\rm i}$ [R$_\odot$] & P$_{\rm f}$ [d] & $a_{\rm f}$ [R$_\odot$] \\
\hline 0.199 & 0&0002  & 0.390 & 0.70 & 4.44 & 0.36  & 2.47 \\ 0.219 &
0&0002 & 0.435 & 0.90 & 5.25 &  0.90 & 4.60 \\ 0.172 & 0&001 & 0.357 &
0.95 &  5.44 & 0.26 & 1.96  \\ 0.183 & 0&001  & 0.391 & 1.02  & 5.71 &
0.43 & 2.79  \\ 0.197 & 0&001 & 0.426  & 1.05 & 5.82 &  0.85 & 4.40 \\
\hline
\end{tabular}
\label{tab.bin}
\end{center}
\end{table}

We want to mention that the assumptions made in our mass loss treatment
give rise to initial He WD models with relatively large hydrogen envelopes,
which favours the occurrence of appreciable hydrogen burning and CNO 
thermonuclear flashes during the WD stage.
This is also true if progenitor stars are more massive than
the 1~M$_\odot$ value assumed in this work. We have checked this assertion
by performing
additional calculations on the basis our mass loss
prescription. However, it is possible that He WDs may also be formed 
in binary systems which have experienced mass loss episodes via
a common-envelope stage, in which case mass transfer would be unstable. 
This could be the case for instance if the accretor is a C/O WD (see, for
instance, Han 1998). This different evolutionary history for the He WD
progenitor could affect
the hydrogen layer thickness and accordingly the He WD cooling rate. 
The treatment of the common envelope stage
is a difficult one and it has not been considered in this paper.

As stated before,  two low metallicity values have  been considered in
this study:  $Z$= 0.001  and $Z$= 0.0002  and we  have computed a  grid of
models for  each of these  values. In the  case of $Z$= 0.0002,  we have
followed the evolution  of He WD models with  stellar masses of 0.199,
0.209, 0.219,  0.225, 0.243, 0.266, 0.280, 0.300  and 0.319 M$_\odot$;
and for the case of $Z$= 0.001 we considered 0.172, 0.183, 0.197, 0.230,
0.244, 0.300, 0.336, 0.380, 0.390,  0.422 and 0.449 M$_\odot$.  All of
these models were evolved from  the end of mass-loss phase, during the
pre-WD evolution,  through the stages of hydrogen  shell flashes (when
they take place) down to very advanced stages of evolution.

\section{Results} \label{sec:results}

\begin{figure*}
\includegraphics[width=17cm,height=20cm]{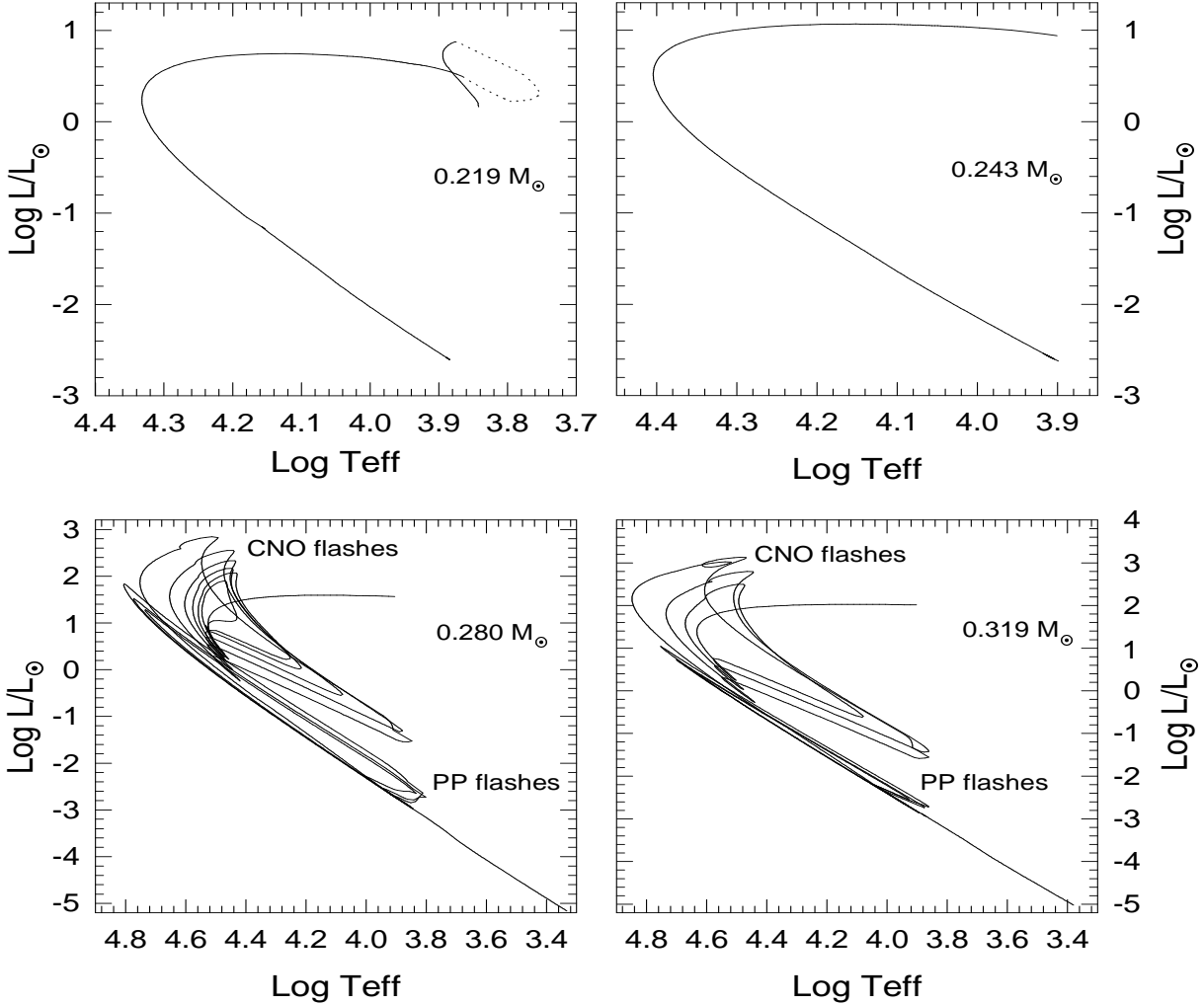}
\vskip -6.2cm
\caption{Evolutionary tracks in  the Hertzsprung-Russell diagram (HRD)
for some  selected masses  of our grid  for $Z$=0.0002. Masses  of
WD models  are shown  in the  plots. For  the 0.219~M$_\odot$
model, pre-WD  evolution is also shown (as  in Fig.~\ref{fig.bin}) and
mass  loss  phase   is  depicted  with  a  dotted   line.   0.219  and
0.243~M$_\odot$ models  do not suffer from  any thermonuclear flashes.
However,  0.280  and  0.319~M$_\odot$  models  undergo  some  hydrogen
flashes.   Soon after  pre-WD evolution  these models  experience some
hydrogen  flashes dominated  by  CNO burning  (loops showing  moderate
redwards  excursions in  the HRD,  labeled  {\it CNO  flashes} in  the
plots). After  these flashes models  cool down but when  T$_{\rm eff}$
reaches approximately 8000K another  sequence of flashes occur, but in
this case  completely dominated by  hydrogen burning via PP  cycle. In
the HRD, these  models evolve ``up'' and ``down'' during
PP-flashes with almost constant radius (labeled as {\it PP flashes} in the
plots).}
\label{fig.tck02}
\end{figure*}

\begin{figure*}
\includegraphics[width=17cm,height=20cm]{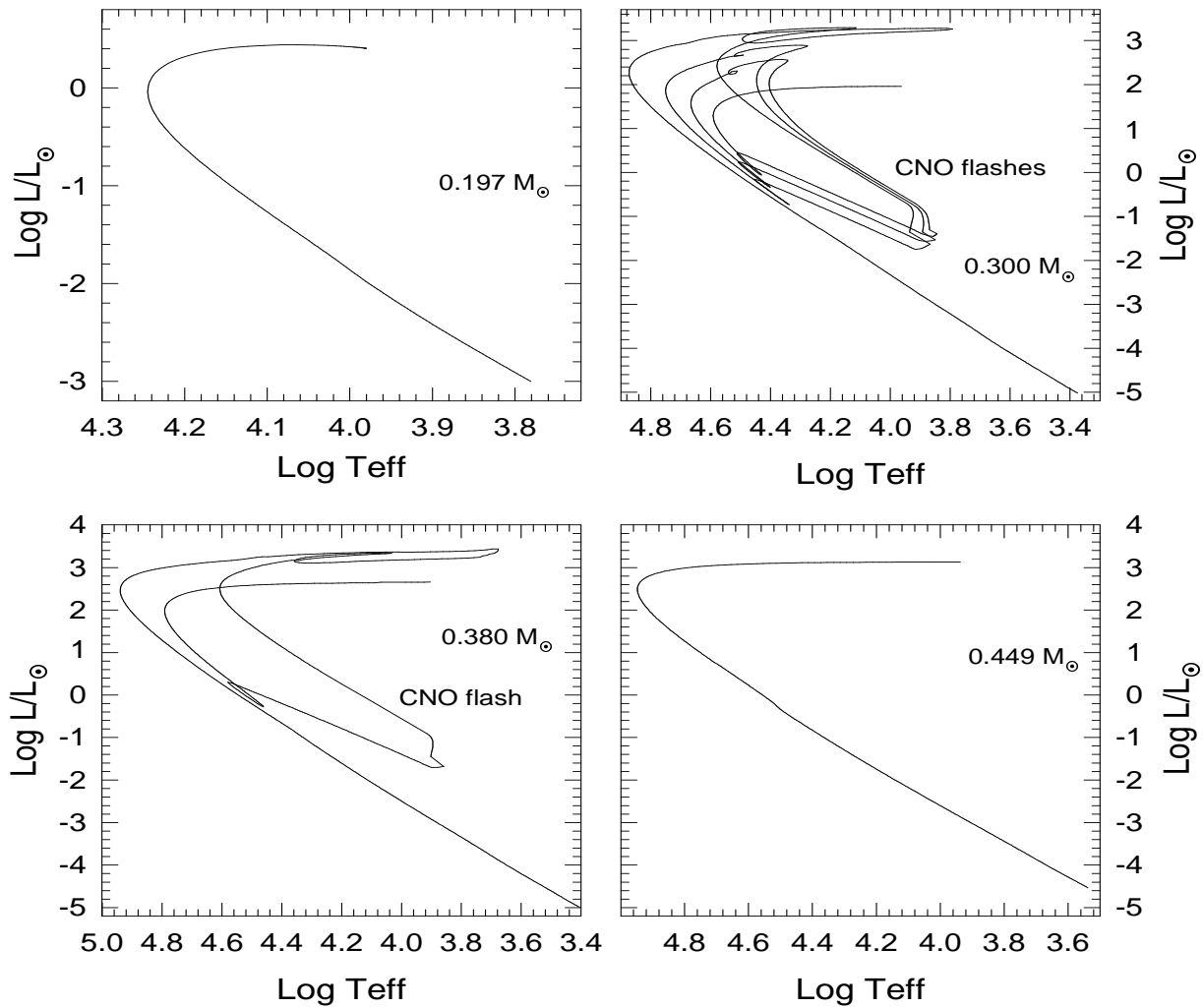}
\vskip -6.2cm
\caption{Evolutionary  tracks in  the Hertzsprung-Russell  diagram for
some selected models from our grid of $Z$= 0.001 HeWDs models. The least
massive model does not suffer from any thermonuclear flashes and cools
very  slowly because  of residual  nuclear burning,  which is  a major
contribution to the energy budget of the star. At intermediate masses,
as represented by our 0.300~M$_\odot$ model, element diffusion induces
the occurrence  of several hydrogen flashes dominated  by CNO burning.
For higher masses the number of CNO flashes dimishes (0.380~M$_\odot$)
and finally for  the highest mass value we find  that no flash occur
at all (0.449~M$_\odot$).}
\label{fig.tck1}
\end{figure*}

Qualitatively, the  evolution of  He WDs with  low metallicity  is not
markedly different from the case  of solar metallicity analysed in our
previous studies (Althaus et al.  2001a). Indeed, we find that objects
with  masses  above  a  certain  threshold  value  experience  several
episodes of hydrogen thermonuclear flashes due to the massive 
hydrogen envelopes resulting from binary evolution. The existence
of such flash episodes affect the further
cooling  history of the star. However, the  low metal  content has  
an important effect on the evolution of HeWDs,  particularly  regarding the 
occurrence and characteristics of  CNO thermonuclear flashes. In broad
outline,  CNO flashes  become less  intense  as the  metal content  is
decreased. This is the main reason why loops in the HRD induced by CNO
flashes are markedly less extended  in the case of lowest metallicity.
This  is  illustrated  in Figs.~\ref{fig.tck02}  and  ~\ref{fig.tck1},
where  we  present  the  evolutionary  tracks of  some  selected  HeWD
sequences  with  an initial  metal  content  $Z$=  0.0002 and  $Z$=  0.001
respectively.
%
%
It is worth mentioning that  the stellar mass threshold  (M$_{\rm  th}$) 
for the occurrence
of   CNO    hydrogen   flashes   depends   on    the   assumed   metal
content. Specifically, for $Z$= 0.001 and $Z$= 0.0002 we find a lower mass
limit for the  occurrence of CNO flashes of M$_{\rm  th} \approx$ 0.22 and 0.26
M$_\odot$, respectively (in the case of solar metallicity  M$_{\rm  th}$
becomes  $\approx$ 0.18 M$_\odot$;  see Althaus  et al.   2001a). With
regard  to the upper  mass boundary,  our most  massive model  with $Z$=
0.0002 and 0.319~M$_\odot$  is fully unstable, whilst for  the case $Z$=
0.001 we  find a CNO hydrogen flash  even for a stellar  mass value as
high  as   0.422~M$_\odot$,  but  not   in  the  sequence   with  $M=$
0.449~M$_\odot$.  We  also want to comment  on the fact  that the mass
range  for  the occurrence  of  hydrogen  flashes  depends on  whether
element diffusion is considered or  not, as we find previously for the
solar metallicity case  (Althaus et al. 2001a).  For  instance, in the
absence  of  diffusion,  only  our 0.280,  0.300  and  0.319~M$_\odot$
sequences undergo thermal instabilities in  the case of $Z$= 0.0002. For
the other  metallicity value,  $Z$= 0.001, models  in the  range between
0.24 and 0.35~M$_\odot$ approximately, suffer from this instabilities.

A  feature worthy  of comment  predicted  by our  calculations is  the
existence of  thermal instabilities in  models with $Z$= 0.0002  at more
advanced  stages  of  evolution   than  those  at  which  CNO  flashes
occur. These  instabilities are  related to unstable  hydrogen burning
via the  proton-proton chains (PP) which dominate  hydrogen burning at
advanced ages. 
The  existence of such PP flashes  can be understood on
the  basis of a  much thicker  hydrogen envelope  characterizing these
models.   This is a  result of  the CNO  burning being  inefficient in
depleting the hydrogen content of  the envelope of the star during the
pre-WD evolution as well as  during CNO flashes. After 
the end of CNO flash episodes, the radial extent of the upper
boundary of the hydrogen
burning shell diminishes as evolution proceeds.  
At the same time,
some hydrogen chemically diffuses downwards to hotter layers.
When the burning shell is thin enough,  nuclear burning becomes
unstable, as predicted by the instability criterion of Kippenhahn
\& Weigert (1990) (see  Driebe et al. 1999 for an application
of this instability criterion in the context of CNO flashes in HeWDs)
\begin{equation}
C^*= C_P\left(1-\frac{4\ \nabla_{ad}\delta}{4\alpha-\frac{R_{shell}}{D}}\right) > 0,
\end{equation}
where $C^*$ and $C_P$ are the gravothermal and constant pressure
specific heats respectively, $\nabla_{ad}$ is the adiabatic gradient, 
$\alpha= \left( \frac{\partial ln \rho}{\partial ln P}  \right)_T$,
$\delta= -\left( \frac{\partial ln \rho}{\partial ln T}  \right)_P$,
$R_{shell}$ is the radial coordinate of the shell and $D$ its radial
extent (defined as the point at which nuclear burning falls below
0.001 of the maximum value). 
We find that this criterion is fulfilled in
our models at the onset of PP flashes at \teff $\approx$
8500K. In particular, this is because of the large 
values reached by the quantity $R_{shell}/D$, which
exceeds  10 at the onset of the flash; while $\alpha$ and $\delta$ are close to 
0.85 and 0.60 respectively.

PP-induced flashes differ from their CNO counterparts in some respects.
For instance, in  most cases PP flashes are less  intense than the CNO
ones and during  flash episodes convective mixing in  the outer layers
of the  star is  less frequent than  for CNO flashes.   An interesting
characteristic of these  flashes is that after a  PP hydrogen flash is
initiated,  the star is  forced to  move in  the HR  diagram following
lines of  almost constant radii close to the cooling  track. Also,
during each of  these PP flashes, the total amount  of hydrogen in the
star is  barely reduced  by nuclear burning  with the result  that the
star in  general will experience  numerous episodes of this  kind.  In
the sequences with $Z$= 0.001 only those models which lack CNO flashes
experience PP hydrogen flashes, but they occur at extremely high ages.
All of our  sequences with $Z$= 0.0002 are characterized by
PP hydrogen flashes even if  they have experienced CNO flashes earlier
in their evolution.  The important point to emphasize here is that the
PP  flashes  are  induced  by  chemical diffusion  that  carries  some
hydrogen downwards deep enough for  the star to ignite hydrogen there.
Except for the  less massive models ($M  \lesssim$  0.25 M$_\odot$), PP
flashes in the sequences with $Z$= 0.0002 take place between
2 and 4 Gyr after the end  of mass loss episodes. In general, the more
massive  the He  WD, the  earlier in  the life  of the  star  these PP
flashes occur.   It is worth  emphasizing again that the  existence of
such PP  flashes in He  WDs is restricted  to the case  of metallicity
much lower than the solar metallicity.  For instance, such flashes are
absent in the  case of models with $Z$= 0.02, and  only our  0.172 and 
0.183 M$_\odot$ models with $Z$= 0.001 suffer from them.  The fact that
essentially  models with  extremely  low metal  content experience  PP
flashes can be understood on the  basis that such models are left with
hydrogen envelopes  so massive that the tail  of hydrogen distribution
chemically diffusing inwards reaches hot enough layers. This increases the
hydrogen burning by PP reactions and leads to a thermal instability.
\begin{figure}
\includegraphics[width=8cm]{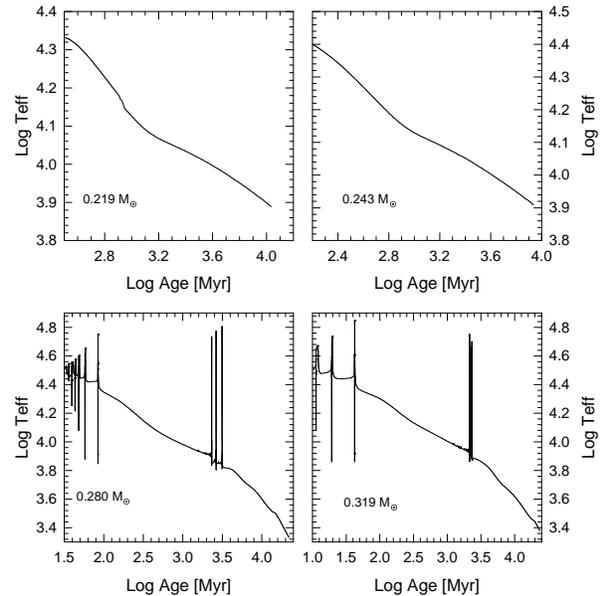}
\vskip -3.5cm
\caption{Effective  temperature as  a  function of  age  for the  same
models  shown in  Fig.~\ref{fig.tck02} ($Z$=  0.0002). For  more massive
models,  spikes   correspond  to  evolutionary   stages  during  flash
episodes. It  is clear that  models which experience  unstable burning
reach very low effective temperatures well within 15 Gyr.}
\label{fig.age02}
\end{figure}
\begin{figure}
\includegraphics[width=8cm]{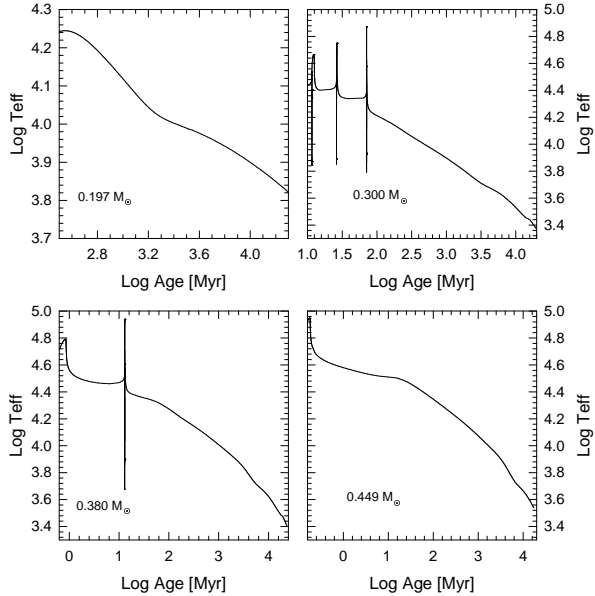}
\vskip -3.5cm
\caption{Effective  temperature as  a  function of  age  for the  same
models shown in Fig.~\ref{fig.tck1}  ($Z$= 0.001).  Spikes correspond to
evolutionary   stages   during    flash   episodes.    As   shown   in
Fig.~\ref{fig.age02}, models  characterized by unstable  burning reach
very  low  effective  temperatures   well  within  15  Gyr.   For  the
0.449~M$_\odot$ sequence  a short evolutionary  time-scale is obtained
even in the absence of thermonuclear flashes.}
\label{fig.age1}
\end{figure}

\begin{figure}
\includegraphics[width=8cm]{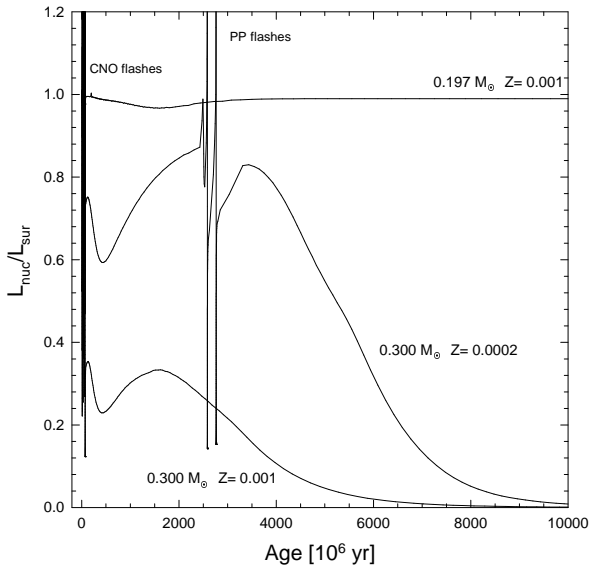}
\vskip -4.1cm
\caption{Hydrogen burning luminosity in terms of surface luminosity as
a function of age for our 0.197  ($Z$= 0.001) and 0.300 ($Z$= 0.001 and 
$Z$=0.0002) He WD models. Note that for the lowest mass shown, the lack of
thermonuclear   flashes  causes  surface   luminosity  to   be  almost
exclusively  supported by  stable hydrogen  burning. In  contrast, for
more  massive  models,  stable  nuclear burning  following  the  flash
episodes contributes only partially to  the energy budget of the star.
Note  also  that as  Z  is  increased,  the contribution  of  residual
hydrogen burning becomes less important.}
\label{fig.nuc}
\end{figure}

In  Figs.~\ref{fig.age02}~and~\ref{fig.age1}  we  show  the  effective
temperature as  a function  of cooling age  for the same  stellar mass
values   as  given   in   Figs.~\ref{fig.tck02}  and   ~\ref{fig.tck1}
respectively.  Note that the lack of CNO thermonuclear flashes in less
massive models  leads to a very  slow rate of evolution,  that is such
models remain relatively  bright even at very large  ages.  This is so
because  stable  hydrogen  shell  burning  via  PP  nuclear  reactions
constitutes  the main  source of  luminosity over  most of  the entire
evolution  of  these  models.   On  the  contrary,  the  existence  of
diffusion-induced CNO thermonuclear flashes leads eventuallyto He 
WD models with
hydrogen  envelope masses  small  enough for  nuclear  burning not  to
appreciably contribute to the luminosity budget of the star. Accordingly,
such models are characterized by short evolutionary time-scales. This age
dichotomy between  models with and without CNO  flashes manifests also
itself in cooling sequences with a solar metal content (Althaus et al.
2001a).  It is worth noting  that such a dichotomy is less accentuated
in  the case  of models  with very  low metal  content. To  place this
assertion on a more quantitative  basis, we mention that in reaching a
value  of   Teff=  8000~K,  models   with  0.219,  0.243,   0.280  and
0.319~M$_\odot$ ($Z$=  0.0002) need  about 10, 8.7,  3.1 and  2.3 Gyr, 
respectively. In the
case of $Z$= 0.001, models  with 0.197, 0.300, 0.380 and 0.449~M$_\odot$
need about 10, 1, 2.2 and 3.1 Gyr, respectively. The dependence of the
cooling history  on the assumed initial metallicity  can be understood
by inspecting Fig.~\ref{fig.nuc} in  which we show the time dependence
of hydrogen burning luminosity  for our 0.300~M$_\odot$ sequences with
$Z$= 0.001 and $Z$= 0.0002. In the  case of $Z$= 0.001, after 1.5 Gyr of the
occurrence of  CNO flashes, stable hydrogen burning  reaches a maximum
contribution  to  surface  luminosity  of 35\%.   Afterwards,  nuclear
burning becomes less important as evolution proceeds. For instance, at
4 Gyr, it only contributes less than 10\%  to the star's luminosity.
In  contrast,  in  the  case  of $Z$=  0.0002,  stable  nuclear  burning
contributes appreciably to surface luminosity even at an age of 6 Gyr.
After  CNO flashes, stable  nuclear burning  contributes   more than
60\% of the energy radiated by  the star within the first 5 Gyr.  This
contribution  drops rather  abruptly during  the subsequent  stages of
evolution,  eventually  reaching  10\%  at  7  Gyr.   With  regard  to
Fig.~\ref{fig.nuc}, let us finally  mention that we have also included
the  0.197~M$_\odot$ sequence ($Z$=  0.001) which  does not  suffer from
flash  episodes.  In  this  case, the  stellar  luminosity is  completely
supported by stable hydrogen  burning along the entire cooling branch,
which is clearly shown in Fig.~\ref{fig.nuc}.

\begin{figure}
\hskip -.2cm \includegraphics[width=8.5cm]{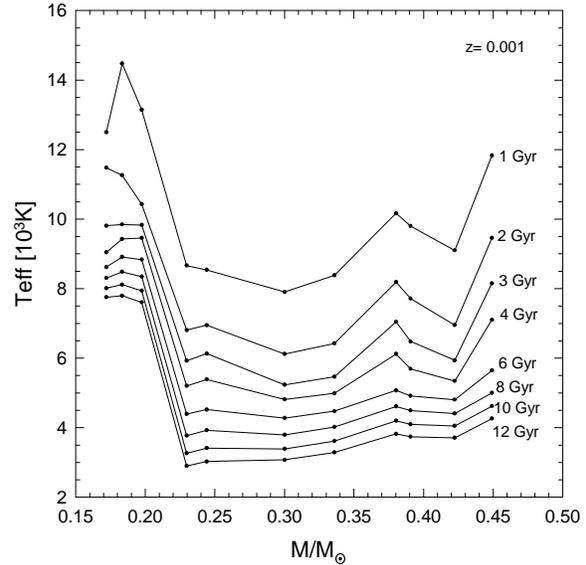}
\vskip -4.7cm
\caption{Isochrones for our He WD  sequences with $Z$= 0.001. The abrupt
drop  shown  by  isochrones  in  the range  0.20-0.22~M$_\odot$  is  a
consequence of the existence of a mass threshold for the occurrence of
CNO flashes. Much shorter cooling ages result if the mass of the model
is above this threshold value. The bump exhibited by the isochrones at
$\approx$ 0.38~M$_\odot$  is related to an increase  in the efficiency
of PP stable nuclear burning at low ages. For more details, see text.}
\label{fig.iso1}
\end{figure}

\begin{figure}
\hskip -.2cm \includegraphics[width=8.5cm]{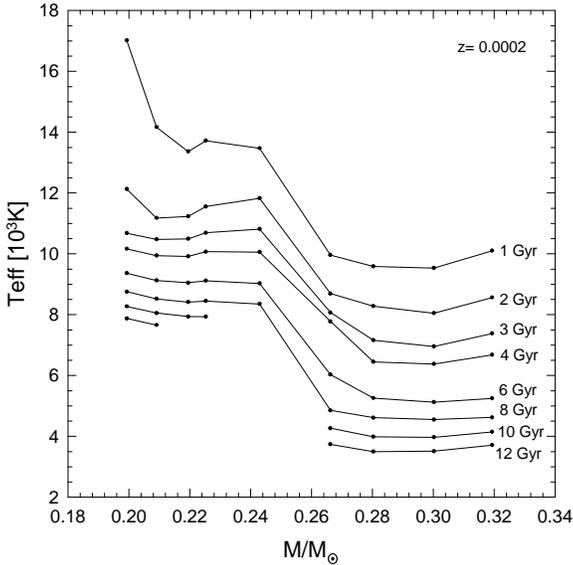}
\vskip -4.7cm
\caption{Same  as Fig.~\ref{fig.iso02}  but  for the  case $Z$=  0.0002.
Here isochrones exhibit a not so  pronounced drop as in the case of $Z$=
0.001  models.    For  models  that  not  suffer   from  CNO  flashes,
calculations were stopped  before PP flashes begin and  thus, some low
mass models have not reached ages as large as 10 Gyr (Because of this
reason, the 10 and 12 Gyr isochrones do not join up).}
\label{fig.iso02}
\end{figure}

In Figs.~\ref{fig.iso1}  and   \ref{fig.iso02}  we  plot  the
isochrones for  some selected  ages in the  mass-effective temperature
plane for  our $Z$= 0.001 and  $Z$= 0.0002 models  respectively.
Note that at the stellar mass threshold, M$_{\rm  th}$, the isochrone plots 
exhibit a pronounced discontinuity. This is not surprising in view of the
age dichotomy discussed earlier. The plot
corresponding to $Z$=  0.001 models resembles that of  He WD models with
an initial solar-like  metal content (Althaus et al.   2001b). This is
particularly true  regarding the presence  of an age dichotomy  at, in
this  case, M$_{\rm  th} \sim  0.22$~M$_\odot$. We  judge  that future
observations of low mass WDs with millisecond pulsar companions formed
in low-metallicity environments could eventually place our predictions
on  solid  observational  grounds\footnote{Tentative support  for  the
existence  of such  an age  dichotomy  is given  by the  field  He  WD
companions  to PSR  B1855+09  and PSR  J1012+5307 millisecond  pulsars
(Althaus et al. 2001ab).}.  Another feature worthy of comment depicted
by Fig.~\ref{fig.iso1}  is the  existence of a  bump in  the isochrone
curves  at  0.380-0.390~M$_\odot$,  particularly  at low  ages.   This
behaviour can be understood on the  basis of the role played by stable
PP  hydrogen burning  after CNO  flash episodes.   Specifically, after
experiencing only one CNO  flash, the 0.380 and 0.390~M$_\odot$ models
are left  with a hydrogen  envelope thin enough  so as to  inhibit the
occurrence of another CNO  flash.  However, envelopes are thick enough
to  mantain an appreciable  amount of  nuclear energy  release through
stable  PP  hydrogen  burning,   thus  slowing  the  cooling  rate  at
intermediate stages  of evolution. Finally,  note the increase  of the
isochrone curves for the most massive sequence, which does not undergo
any thermonuclear flash, thus resulting in comparatively large cooling
ages.   The  situation  for  the   case  of  $Z$=  0.0002  sequences  is
qualitatively similar, though the age dichotomy is less noticeable and
it is  located at M$_{\rm th} \sim  0.25-0.26$~M$_\odot$. As explained
earlier, this is a result of the fact that, in this case, during flash
episodes  CNO  cycle reactions  are  less  efficient  than for  higher
metallicities  and  thus,  hydrogen  envelope is  not  so  drastically
consumed.

\begin{figure*}
\includegraphics[width=15cm,height=17cm]{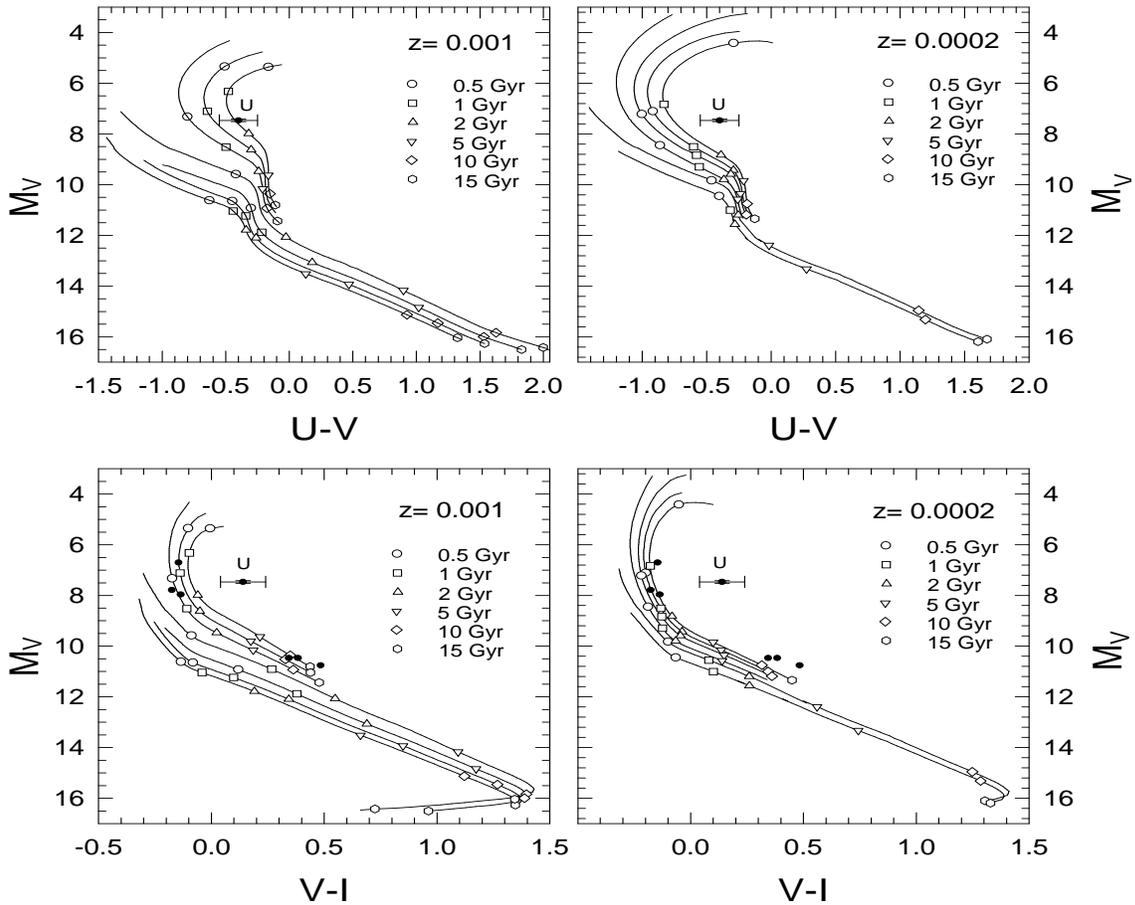}
\vskip -4.5cm
\caption{Absolute  visual magnitude  in  terms of  the colour  indices
$U-V$ (upper panels) and $V-I$  (bottom panels) for He WDs models with
stellar  masses (from right to left)  0.172,  0.183,   0.197,  0.230,  
0.300,   0.380  and
0.449~M$_\odot$ for  $Z$= 0.001, and  0.199, 0.209, 0.219,  0.243, 0.266
and 0.319~M$_\odot$  for $Z$= 0.0002.   Symbols along the  curves denote
some selected ages.  In addition, the observational data for the He WD
in 47 Tucanae  (Edmonds et al. 2001; denoted by ``U'')  and the six He
WD candidates  in NGC  6397 reported by  Taylor et al.   (2001; filled
circles) are included in the figure.}
\label{fig.col}
\end{figure*}

The evolution of our He  WD models in the colour-magnitude diagrams is
illustrated in Fig.~\ref{fig.col}, which shows the run of the absolute
visual  magnitude $M_V$ for  our models  as a  function of  the colour
indices $U-V$  and $V-I$.  Note  that models which have  suffered from
CNO thermonuclear flashes show a  turn-off in the $V-I$ colour at $M_V
\approx$ 16 and  $V-I \approx$ 1.4. This turn-off,  which results from
the strong CIA  opacity by molecular hydrogen at  low temperatures, is
reached well within  15 Gyr, mostly by He WDs  with $Z$= 0.001.  After
reaching the turn-off, the emergent spectrum becomes bluer and remains
brighter than  $M_V \approx $  17 with subsequent cooling.   A similar
trend has been also found for the case of solar metallicity (Serenelli
et al.   2001), though  in that situation  He WDs with  stellar masses
even as low as $\approx$ 0.19~M$_\odot$ reach high magnitude values.

In Fig.~\ref{fig.col}  we have also included  the recent observational
data of Taylor  et al (2001) and  Edmonds et al. (2001) for  the He WD
candidates  in  the  globular   clusters  NGC  6397  and  47  Tucanae,
respectively.   For  NGC  6397  we  have adopted  a  distance  modulus
(m-M)$_{\rm V}$= 12.29  and E(B-V)= 0.18 (Cool et  al.  1998).  In the
case of 47 Tucanae, (m-M)$_{\rm  V}$= 13.39 and E(B-V)= 0.055 (Zoccali
et  al.  2001)  were  used  to transform  to  absolute magnitudes  and
to deredden colour indices.  In both cases, exctintion laws from Holtzman
et al.  (1995) were applied. Let us consider first the HeWD candidates
in  NGC 6397. Given  the metal  content for  NGC 6397  ([Fe/H= -1.82];
Anthony-Twarog  \& Twarog 2000),  then our  $Z$= 0.0002  sequences are
appropriate for a direct comparison with observations.  Note that the
six  HeWD candidates  in NGC  6397 can  be split  into two  groups. In
particular, for  the three brightest  stars of the  observed sequence,
our evolutionary models fit the observational data quite well. We find
that such He WDs would be  characterized by stellar mass values in the
range  0.2-0.22~M$_\odot$,  which is  below  the  mass  value for  the
occurrence of CNO flashes.  With  regard to age determination, we find
ages  between 0.5-1.5~Gyr  for  these stars.  By  contrast, the  three
dimmest stars  appear to have larger  radii even as  compared with our
least  massive model ($\approx  0.2$M$_\odot$), 
as it is clear from the lower right panel of Fig.~~\ref{fig.col}.
 We  want to  point out
that,  based  on our  mass loss treatment, we find a
minimum mass value for the resulting He core remnant which is very close
to  0.2 M$_\odot$ for progenitors with $Z$= 0.0002.
We have  checked this by
performing additional pre-WD calculations for different initial binary
configurations. In  this context,  Burderi, D'Antona \&  Burgay (2002)
have also found that their  evolutionary models have smaller radii than
those required to  fit the observational data. In  addition, they also
find a  minimum mass  limit of $\approx  0.2$M$_\odot$ for  HeWDs with
very low metallicity progenitors. However, it is worth noting that our
models  are  somewhat less  compact  than  those  presented by  Burderi  et
al. (2002).  We judge that this  difference could be a  result of the
fact  that  in our  calculations  we  have  included chemical  element
diffusion which  leads to pure  hydrogen envelopes and thus  to models
with larger radii (Althaus et al.  2001a), thus reducing the discrepancy
with observations. Another point worthy of 
comment in connection with this second group of HeWD candidates is related
to their ages. Indeed, our least massive sequence yields cooling ages 
between 10 and 15 Gyr for these stars.\footnote{We  stress again 
that such large  ages are expected because our less massive models are 
characterized by mass values well below the M$_{\rm th}$ value for the 
occurrence of CNO thermonuclear flashes.} If we take into account the time
elapsed during the pre-WD evolution (which is about 6-7~Gyr for a 1~M$_\odot$
progenitor), then the age of these stars would be in the range 16-21~Gyr, 
larger than the age of the cluster, which is $12\pm0.8$~Gyr (Anthony-Twarog  
\& Twarog 2000). 
This suggests that the progenitors
of these HeWD candidates should have been more massive than 1~M$_\odot$ with, 
therefore, shorter pre-WD evolutionary ages. To explore this 
possibility, we have performed additional calculations with our mass loss
prescription. For instance, we found that
starting from a 1.4~M$_\odot$ progenitor with a 1.4~M$_\odot$ companion 
and an initial period of 0.5 days, the pre-WD evolution lasts for about
3~Gyr and the mass of the WD remnant is 0.195~M$_\odot$. This would 
help to alleviate the age discrepancy mentioned above. 
In this regard, we believe that a more comprehensive exploration of the
binary nature leading to the formation of HeWDs, particularly for progenitors
with very low metal content, is worth carrying out.

With regard to the He WD candidate in the globular cluster 47 Tucanae,
we find  that our evolutionary  models with $Z$= 0.001,  a metallicity
value  not very  different from  that of  47 Tucanae  ([Fe/H= -0.76]),
yield  a good  agreement with  observations  for the  $U-V$ index.  In
particular, we  derive a  stellar mass value  of $\approx$
0.17~M$_\odot$  and a  cooling  age of  about  1.5 Gyr for the WD.   
This age  is
consistent with  the 2 Gyr corresponding  to the spin-down  age of the
millisecond  pulsar companion  to the  WD  (Edmonds et  al.  2001  and
referenced cited  therein). Note  that, as far  as the $V-I$  index is
concerned, we find no agreement  with observational data. In fact, our
evolutionary models are located too far  to the blue. A
lower stellar  mass value (M $\approx$ 0.15-0.16~M$_\odot$)  than our
lowest mass sequence  would be needed to match  the observed $V-I$. In
the case of  $Z$= 0.0002 and for the stellar  masses we considered, we
find no agreement with observations for neither of the metallicities.

Finally, we list in Table~\ref{tab.z1} and~\ref{tab.z2}  some
colour indices for He WD models with, respectively, $Z$= 0.0002 and 
$Z$= 0.001 at selected effective temperature values. 
We also list the surface gravity ($g$), the age (in Gyr), the
bolometric correction ($BC$) and the absolute visual magnitude ($M_V$).

\section{Conclusions} \label{sec:conclusion}

In  this study  we have  explored the  evolution of  helium-core white
dwarf (He  WD) stars with progenitors having  much lower metallicities
than the  solar metallicity usually  assumed in the modeling  of these
stars.    The  models   presented   here  are   appropriate  for   the
interpretation of  recent and future  observations of low-mass  WDs in
globular clusters.  Specifically, two low metallicity values have been
considered: $Z$= 0.001  and $Z$= 0.0002.  In the  case of $Z$= 0.0002,
we have followed the evolution of  He WD models with stellar masses of
0.199,  0.209, 0.219,  0.225,  0.243, 0.266,  0.280,  0.300 and  0.319
M$_\odot$; and for the case  of $Z$= 0.001 we considered 0.172, 0.183,
0.197,  0.230, 0.244,  0.300,  0.336, 0.380,  0.390,  0.422 and  0.449
M$_\odot$.  All of these models were evolved from the end of mass-loss
phase down to very  advanced phases of evolution.  The binary
nature of  our He  WD models has  been simulated by  abstracting mass
from a  1 M$_\odot$ model at  appropriate stages of  its evolution off
the main  sequence.  Specifically, to obtain  physically sound initial
low-mass He WD  models, mass transfer rates  were derived by  imposing that the
stellar radius remains close to the radius of the Roche lobe.  A fully
non-conservative approach was assumed and the loss of angular momentum
caused by gravitational wave emission as well as magnetic stellar wind
braking have been  taken into account.

The  evolution   of  our  He  WD   models  has  been   computed  in  a
self-consistent  way with  the predictions  of  time-dependent element
diffusion and nuclear burning. A non-gray treatment for the atmosphere
that considered  the energy transfer  by radiation and  convection has
been employed to derive the  outer boundary conditions of our evolving
models. Models  atmosphere are  based on a  detailed treatment  of the
microphysics entering the WD atmospheres, such as non-ideal effects in
the equation of state (as  given by the occupation formalism of Hummer
\& Mihalas 1988) and the  inclusion of hydrogen line (from the Balmer,
Lyman  and  Paschen  series)  and  pseudo-continuum  opacities.  Also,
up-to-date  collision-induced absorption  data (Rohrmann  et  al. 2002) 
were  incorporated  in the  computations.  Such  a
detailed description allowed  us to provide a grid  of accurate colour
indices and magnitudes at both early and advanced evolutionary stages, 
strongly improving  previous efforts (Serenelli et al. 2001).

A  feature worthy  of comment  predicted  by our  calculations is  the
existence  of thermonuclear  flash  episodes  for most  of  our He  WD
sequences.  In part, this  is  a  result  of  including  in  our
computations the  various chemical  diffusion processes.   We find
that the lower the metallicity  $Z$, the larger the minumum stellar mass 
for the  occurrence of hydrogen  thermonuclear flashes induced  by the
CNO cycle reactions.  Specifically, for  $Z$= 0.001 and $Z$= 0.0002 we
find  a lower  mass  limit for  the  existence of  CNO  flashes of  $M
\approx$ 0.22  and 0.26 M$_\odot$,  respectively (in the case  of $Z$=
0.02 such a limit becomes $\approx$ 0.18 M$_\odot$; see Althaus et al.
2001a).  In  addition, CNO  flashes become less  intense as  the metal
content of the star is decreased and the mass range for the occurrence
of such  instabilities depends strongly  on the inclusion  of chemical
diffusion processes  in evolutionary calculations.  We  also find that
the   existence  of   a   mass  threshold   for   the  occurrence   of
diffusion-induced CNO flashes leads  to an age dichotomy (particularly
for  $Z$=   0.001)  between  He   WD  models  with  and   without  CNO
thermonuclear   flashes:  He   WDs   which  do   not  experience   CNO
thermonuclear flashes  evolve very  slowly, so they  remain relatively
bright even  at very large ages,  whilst those which  suffer from such
flashes  are  characterized  by  a  fast cooling,  reaching  very  low
effective temperature  stages within cooling  times less than  15 Gyr.
Such an age  dichotomy is translated into distinctive  features in the
isochrone plots which could  eventually be compared with observational
expectations.
It is worth mentioning that  during the short-lived, 
CNO diffusion-induced flashes, an 
appreciable amount of hydrogen is burnt, ultimately implying that the WD
is left with a relatively thin hydrogen envelope which prevents stable
hydrogen burning from being an important energy source at late cooling stages.
As a result, the star has a much lower amount of available energy, which 
implies much shorter evolutionary time scales as compared with the situation 
when CNO flashes are absent.

Another finding of  this work is related to the fact  that some of our
He  WD models  experience  several episodes  of thermal  instabilities
related  to  unstable  hydrogen  burning via  the  proton-proton  (PP)
nuclear  reactions.   These  PP  flashes,  which take  place  at  more
advanced  stages of  evolution than  those  at which  the CNO  flashes
occur, are  experienced by  all of our  sequences with $Z$=  0.0002.  In
addition, our  0.172 and 0.183  M$_\odot$ models with $Z$=  0.001 suffer
from   PP flashes,  but at exceedingly  high ages.  We find
that PP  thermal instabilities are triggered by  chemical diffusion that
carries some  hydrogen downwards  deep enough for  the star  to ignite
hydrogen there.  Except for the less massive models ($M \lesssim$ 0.25
M$_\odot$), PP flashes take place between 2 and 4 Gyr after the end of
mass loss  episodes, and in general,  the more massive the  He WD, the
earlier in the star life they occur.

The  evolution of  our He  WD  models has  also been  analysed in  the
colour-magnitude  diagrams. We  find that  models which  have suffered
from  CNO flashes  exhibit  a turn-off  in  most of  their colours  at
M$_{\rm V} \approx$  16. This turn-off, which results  from the strong
CIA opacity by molecular hydrogen at low temperatures, is reached well
within  15  Gyr,  mostly by  He  WDs  with  $Z$= 0.001.  Finally,  the
predictions of our models  for the colour-magnitude diagrams have been
compared with  recent observational  data of He  WD candidates  in the
globular clusters  NGC 6397  and 47 Tucanae  (Taylor et al.   2001 and
Edmonds et al.  2001, respectively).  In this connection, we find that
the  three  brightest HeWD  candidates in  NGC  6397  can indeed  be
identified  as  HeWDs   characterized  by  stellar  mass  values  of
0.20-0.22~M$_\odot$ (which is below  the mass value for the occurrence
of CNO flashes) and ages ranging  from 0.5 to 1.5 Gyr. However, in the
case of  the three  dimmest candidates, the agreement with observational data
is not so evident as in the case of the brightest objects. Indeed, our  models
appear to be  more massive than required by  observations. Specifically, 
our mass-loss  treatment gives  rise to  a minimum value of
$\sim$0.2~M$_\odot$ an  HeWD may have  from progenitors with initially 
$Z$= 0.0002. However, it is worth noting that a $\Delta(V-I)\approx 0.1$
would allow a very good agreement between theoretical predictions and 
observations. In this sense, it is remarkable that the quoted 
uncertainty in the case of the HeWD candidate
in 47 Tucanae is very close to this value (see Fig.10)\footnote{It is 
worth mentioning that Townsley \& Bildsten (2002) have 
recently suggested that the three dimmest HeWD candidates in NGC 6397 
could be cataclismic variables involving a hot C-O WD and a low-mass 
main sequence companion.}.
Finally
the He WD in 47
Tucanae  is particularly  relevant because  the spin-down  age  of its
millisecond pulsar  companion yields a WD  age estimate. Specifically,
the pulsar age of 2 Gyr (see Edmonds et al. 2001) is in agreement with
the prediction  of our $Z$= 0.001  He WD sequences.   In addition, the
observational data  for the M$_{\rm  V}$ and $U-V$ is  consistent with
our 0.17~M$_\odot$ sequence.

Complete tables containing the results of the present calculations are
available  at  http://www.fcaglp.unlp.edu.ar/evolgroup/  or  upon
request to the authors at their e-mail addresses. 

\begin{table*}
\centering
\begin{scriptsize}
\begin{minipage}{135mm}
\caption{Selected  stages for  0.199, 0.243  and 0.30~M$_\odot$  He WD
models for metallicity $Z$= 0.0002}
\begin{tabular}{@{}ccccccccccccc@{}}
\hline $M_*/{\rm M_{\odot}}$  & \teff & $Log (g)$ &  $Age$ (Gyr) & U-B
&B-V & V-R & V-K & R-I & J-H & H-K & BC & M$_V$ \\ \hline 0.199 & 8920
& 4.3218 & 0.295 & -0.09 & 0.10 &  0.04 & 0.12 & 0.06 & 0.07 & -0.09 &
-0.09 & 4.41 \\ '' & 10780 &  4.5281 & 0.494 & -0.24 & -0.05 & -0.03 &
-0.20 & -0.02 &  0.02 & -0.10 & -0.40 & 4.42 \\ ''  & 14220 & 5.0085 &
0.683 & -0.52 & -0.13 & -0.07 &  -0.45 & -0.07 & -0.02 & -0.12 & -1.08
& 5.09 \\ '' & 16150 & 5.3500  & 0.791 & -0.64 & -0.15 & -0.08 & -0.54
& -0.09 & -0.04 & -0.13 & -1.41  & 5.72 \\ '' & 16880 & 5.5465 & 0.861
& -0.68 & -0.16 & -0.08 & -0.57 & -0.10 & -0.04 & -0.13 & -1.52 & 6.14
\\ '' & 17110 & 5.6901 & 0.924 & -0.69 & -0.15 & -0.08 & -0.58 & -0.10
& -0.04 & -0.13 & -1.56 & 6.47  \\'' & 16810 & 5.9008 & 1.054 & -0.68 & 
-0.14 & -0.08 & -0.56 & -0.10 & -0.04
& -0.13 & -1.51 & 7.03 \\ ''  & 16430 & 5.9817 & 1.124 & -0.66 & -0.13
& -0.08 & -0.55 & -0.09 & -0.04 & -0.13 & -1.45 & 7.27 \\ '' & 15960 &
6.0522 & 1.201 & -0.64 & -0.12 & -0.08 & -0.53 & -0.09 & -0.03 & -0.13
& -1.37 & 7.49 \\ '' & 15430  & 6.1141 & 1.283 & -0.61 & -0.11 & -0.07
& -0.50 & -0.08 & -0.03 & -0.12  & -1.28 & 7.71 \\ '' & 14870 & 6.1704
& 1.375  & -0.58 &  -0.09 & -0.07  & -0.47 & -0.08  & -0.02 &  -0.12 &
-1.19 & 7.91 \\ '' & 14280 &  6.2219 & 1.478 & -0.54 & -0.08 & -0.07 &
-0.45 & -0.07 & -0.02 & -0.12 &  -1.08 & 8.11 \\ '' & 13700 & 6.2699 &
1.594 & -0.51 & -0.06 & -0.06 &  -0.41 & -0.06 & -0.01 & -0.12 & -0.97
& 8.30 \\ '' & 13110 & 6.3151  & 1.725 & -0.47 & -0.04 & -0.06 & -0.38
& -0.06 & -0.01 & -0.12 & -0.86  & 8.49 \\ '' & 12530 & 6.3582 & 1.876
& -0.43 & -0.01 & -0.05 & -0.34  & -0.05 & 0.00 & -0.12 & -0.74 & 8.68
\\ '' & 11960 & 6.4003 & 2.058  & -0.39 & 0.02 & -0.04 & -0.29 & -0.04
& 0.01 & -0.11 & -0.63 & 8.88 \\ '' & 11430 & 6.4419 & 2.295 & -0.36 &
0.05 &  -0.03 & -0.24 & -0.02  & 0.02 & -0.11  & -0.53 & 9.07  \\ '' &
10930 & 6.4858 & 2.669 & -0.35 & 0.08 & -0.02 & -0.18 & -0.01 & 0.03 &
-0.11 & -0.43 &  9.28 \\ '' & 10470 & 6.5314 & 3.356  & -0.35 & 0.11 &
0.00 &  -0.12 & 0.01  & 0.03  & -0.11 &  -0.34 & 9.49  \\ '' &  9990 &
6.5726 & 4.351 &  -0.37 & 0.15 & 0.03 & -0.04 & 0.04  & 0.05 & -0.10 &
-0.26 &  9.72 \\ '' & 9510  & 6.6103 & 5.577  & -0.39 & 0.18  & 0.06 &
0.08 &  0.07 & 0.06  & -0.10 & -0.21  & 9.98 \\  '' & 9040 &  6.6448 &
6.985 &  -0.42 & 0.22 & 0.10  & 0.21 & 0.10  & 0.08 & -0.09  & -0.17 &
10.24 \\  '' & 8570 & 6.6772  & 8.658 & -0.45  & 0.26 & 0.13  & 0.34 &
0.13 & 0.10 &  -0.08 & -0.14 & 10.53 \\ '' & 8110  & 6.7069 & 10.622 &
-0.47 & 0.29 & 0.17 & 0.49 & 0.17 & 0.13 & -0.07 & -0.13 & 10.82 \\'' 
&  7250 & 6.7575 & 15.738 & -0.48 & 0.36 &
0.23 & 0.81 & 0.23 & 0.18 &  -0.04 & -0.11 & 11.42 \\ \\ 0.243 & 15050
& 4.4210  & 0.131 & -0.59  & -0.18 & -0.08  & -0.51 & -0.09  & -0.03 &
-0.12 & -1.21 & 3.29 \\ '' &  21020 & 5.0959 & 0.161 & -0.84 & -0.23 &
-0.10 & -0.72 &  -0.13 & -0.07 & -0.14 & -2.10 & 4.42  \\ '' & 23930 &
5.4719 & 0.178 & -0.93 & -0.24 & -0.11 & -0.80 & -0.14 & -0.08 & -0.15
& -2.44 & 5.13 \\ '' & 25070  & 5.7041 & 0.190 & -0.95 & -0.24 & -0.12
& -0.82 & -0.15 & -0.08 & -0.15  & -2.56 & 5.63 \\ '' & 25360 & 5.8768
& 0.201  & -0.96 &  -0.24 & -0.12  & -0.83 & -0.15  & -0.09 &  -0.16 &
-2.59 & 6.04 \\'' & 24650 & 6.1290 &
0.229 & -0.94 & -0.23 & -0.12 &  -0.81 & -0.14 & -0.08 & -0.15 & -2.51
& 6.72 \\ '' & 23920 & 6.2282  & 0.250 & -0.93 & -0.22 & -0.11 & -0.79
& -0.14 & -0.08 & -0.15 & -2.43  & 7.02 \\ '' & 23020 & 6.3141 & 0.277
& -0.91 & -0.21 & -0.11 & -0.77 & -0.14 & -0.07 & -0.15 & -2.33 & 7.30
\\ '' & 22000 & 6.3880 & 0.310 & -0.88 & -0.19 & -0.10 & -0.74 & -0.13
& -0.07 & -0.15 & -2.21 & 7.56  \\ '' & 20900 & 6.4526 & 0.350 & -0.85
& -0.18 & -0.10 & -0.71 & -0.12 & -0.06 & -0.14 & -2.08 & 7.81 \\ '' &
19780 & 6.5089 & 0.396 & -0.81 & -0.16 & -0.10 & -0.67 & -0.12 & -0.06
& -0.14 & -1.93 & 8.05 \\ ''  & 18650 & 6.5594 & 0.451 & -0.77 & -0.14
& -0.09 & -0.64 & -0.11 & -0.05 & -0.14 & -1.78 & 8.27 \\ '' & 17540 &
6.6058 & 0.517 & -0.72 & -0.12 & -0.09 & -0.59 & -0.10 & -0.04 & -0.13
& -1.61 & 8.49 \\ '' & 16460  & 6.6489 & 0.596 & -0.68 & -0.10 & -0.08
& -0.55 & -0.09 & -0.04 & -0.13  & -1.44 & 8.71 \\ '' & 15420 & 6.6900
& 0.695  & -0.63 &  -0.08 & -0.07  & -0.51 & -0.08  & -0.03 &  -0.13 &
-1.27 & 8.92 \\ '' & 14450 &  6.7296 & 0.826 & -0.57 & -0.05 & -0.07 &
-0.46 & -0.07 & -0.02 & -0.12 &  -1.10 & 9.13 \\ '' & 13540 & 6.7702 &
1.031 & -0.52 & -0.02 & -0.06 &  -0.41 & -0.06 & -0.01 & -0.12 & -0.93
& 9.34 \\ '' & 12700 & 6.8121 & 1.418 & -0.46 & 0.01 & -0.05 & -0.35 &
-0.05 & 0.00 &  -0.12 & -0.77 & 9.57 \\ '' & 11870  & 6.8494 & 1.999 &
-0.42 & 0.05 & -0.04 & -0.28 &  -0.03 & 0.01 & -0.11 & -0.61 & 9.79 \\
'' & 11060  & 6.8826 & 2.771 & -0.41  & 0.09 & -0.01 &  -0.20 & 0.00 &
0.03 & -0.11 & -0.45 & 10.02 \\  '' & 10290 & 6.9114 & 3.712 & -0.42 &
0.14 & 0.02 & -0.08 & 0.03 & 0.04 & -0.11 & -0.31 & 10.27 \\ '' & 9560
& 6.9384 & 4.897 & -0.44 & 0.18 &  0.07 & 0.09 & 0.07 & 0.06 & -0.10 &
-0.23 & 10.58  \\'' & 8240 & 6.9865 &
8.308 &  -0.49 & 0.26 & 0.16  & 0.46 & 0.16  & 0.12 & -0.07  & -0.14 &
11.25 \\\\ 
0.300  & 24990 & 6.9774 &  0.065 & -0.97 & -0.21  & -0.12 &
-0.82 & -0.15 & -0.08 & -0.16 &  -2.53 & 8.57 \\ '' & 23390 & 6.9941 &
0.067 & -0.93 & -0.19 & -0.11 &  -0.78 & -0.14 & -0.07 & -0.15 & -2.36
& 8.73 \\ '' & 21800 & 7.0043  & 0.076 & -0.89 & -0.17 & -0.10 & -0.74
& -0.13 & -0.07 & -0.15 & -2.18  & 8.88 \\ '' & 20050 & 7.0178 & 0.105
& -0.83 & -0.15 & -0.10 & -0.68 & -0.12 & -0.06 & -0.14 & -1.96 & 9.06
\\ '' & 18730 & 7.0331 & 0.133 & -0.79 & -0.13 & -0.09 & -0.64 & -0.11
& -0.05 & -0.14 & -1.78 & 9.21  \\ '' & 17510 & 7.0495 & 0.162 & -0.74
& -0.10 & -0.09 & -0.60 & -0.10 & -0.04 & -0.13 & -1.60 & 9.36 \\ '' &
16160 & 7.0699 & 0.199 & -0.68 & -0.08 & -0.08 & -0.54 & -0.09 & -0.03
& -0.13 & -1.39 & 9.55 \\ ''  & 14930 & 7.0903 & 0.241 & -0.62 & -0.05
& -0.07 & -0.48 & -0.07 & -0.02 & -0.12 & -1.18 & 9.73 \\ '' & 13770 &
7.1113 & 0.293 & -0.55 & -0.01 & -0.06 & -0.42 & -0.06 & -0.01 & -0.12
& -0.97 & 9.93 \\ '' & 12870 & 7.1293 & 0.350 & -0.50 & 0.03 & -0.05 &
-0.36 & -0.05 & 0.00 & -0.12 &  -0.80 & 10.10 \\ '' & 12050 & 7.1470 &
0.424 & -0.46 & 0.06 & -0.04 &  -0.29 & -0.03 & 0.01 & -0.12 & -0.64 &
10.27 \\ '' & 11110 & 7.1691 &  0.558 & -0.45 & 0.10 & -0.01 & -0.20 &
0.00 & 0.02 &  -0.11 & -0.45 & 10.48 \\ '' & 10270  & 7.1897 & 0.749 &
-0.46 & 0.15 &  0.03 & -0.06 & 0.04 & 0.04 & -0.11  & -0.32 & 10.75 \\
'' & 9470 & 7.2094 & 1.024 & -0.48  & 0.19 & 0.08 & 0.12 & 0.08 & 0.07
& -0.09 & -0.23 & 11.06 \\ '' & 8870 & 7.2239 & 1.336 & -0.50 & 0.22 &
0.12 & 0.28  & 0.12 & 0.09  & -0.08 & -0.19 &  11.34 \\ A ''  & 8290 &
7.2368 & 1.766  & -0.51 & 0.26 & 0.16  & 0.46 & 0.16 &  0.12 & -0.07 &
-0.15 & 11.63 \\ B '' & 7800 &  7.2634 & 2.776 & -0.51 & 0.30 & 0.19 &
0.61 &  0.20 & 0.15 & -0.06  & -0.13 & 11.94  \\ '' & 7150  & 7.2688 &
2.796 &  -0.48 & 0.37 & 0.24  & 0.85 & 0.25  & 0.19 & -0.03  & -0.12 &
12.32 \\  '' & 6570 & 7.2731  & 3.751 & -0.42  & 0.44 & 0.29  & 1.09 &
0.30 & 0.23  & -0.01 & -0.11 & 12.69  \\ '' & 6060 &  7.2852 & 4.394 &
-0.33 & 0.53 & 0.35 & 1.33 & 0.35  & 0.27 & 0.01 & -0.12 & 13.08 \\'' & 
5210  & 7.3257 & 5.748 &  -0.11 & 0.72 &
0.47 & 1.82 & 0.47 & 0.32 & 0.07 & -0.21 & 13.93 \\ '' & 4840 & 7.3486
& 6.969 &  0.00 & 0.82 &  0.53 & 2.07 & 0.53  & 0.35 & 0.09  & -0.31 &
14.40 \\ '' & 4470 & 7.3622 & 8.269 & 0.12 & 0.92 & 0.59 & 2.23 & 0.59
& 0.33 & 0.04 &  -0.42 & 14.89 \\ '' & 4110 & 7.3696  & 9.479 & 0.22 &
1.01 & 0.65 & 2.12 & 0.65 & 0.12  & -0.02 & -0.46 & 15.32 \\ '' & 3770
& 7.3743 & 10.729 & 0.29 & 1.07 & 0.69 & 1.67 & 0.68 & -0.14 & -0.14 &
-0.41 & 15.65  \\ '' & 3460 & 7.3785  & 12.260 & 0.36 &  1.13 & 0.72 &
1.12 & 0.68  & -0.30 & -0.28 & -0.31  & 15.93 \\ '' &  3180 & 7.3835 &
15.016 & 0.44  & 1.19 & 0.72 & 0.59  & 0.60 & -0.37 &  -0.36 & -0.16 &
16.16 \\  '' & 2920 & 7.3858  & 16.606 & 0.51  & 1.24 & 0.69  & 0.05 &
0.43 & -0.38 & -0.47 & 0.01 & 16.37 \\ \hline
\end{tabular}
\label{tab.z1}

{\small Ages are counted from the end of mass transfer. Letters A and B
denote the age  interval during which the 0.30  M$_\odot$ model return
several  times to  high  \teff values  as  a result  from PP  hydrogen
flashes.}
\end{minipage}
\end{scriptsize}
\end{table*}

\begin{table*}
\centering
\begin{scriptsize}
\begin{minipage}{135mm}
\caption{Selected stages  for 0.172, 0.230, 0.336  and 0.449 M$_\odot$
He WD models for metallicity $Z$= 0.001}
\begin{tabular}{@{}ccccccccccccc@{}}
\hline $M_*/{\rm M_{\odot}}$  & \teff & $Log (g)$ &  $Age$ (Gyr) & U-B
&B-V & V-R & V-K & R-I & J-H & H-K & BC & M$_V$ \\ \hline

  0.172 & 9390 & 4.6630 & 0.349 &  -0.13 & 0.07 & 0.02 & 0.02 & 0.03 &
  0.05 & -0.09 & -0.16 & 5.27 \\ '' & 12420 & 5.2994 & 0.973 & -0.40 &
  -0.07 & -0.05 & -0.33 & -0.05 &  0.00 & -0.11 & -0.74 & 6.22 \\ '' &
  12760 &  5.4971 & 1.173 &  -0.42 & -0.07 &  -0.05 & -0.35  & -0.05 &
  0.00 & -0.11 & -0.81 & 6.67 \\ '' & 12700 & 5.6392 & 1.356 & -0.42 &
  -0.06 & -0.05 & -0.35 & -0.05 &  0.00 & -0.11 & -0.79 & 7.03 \\ '' &
  12410 &  5.7509 & 1.544 &  -0.40 & -0.05 &  -0.05 & -0.33  & -0.05 &
  0.00 & -0.11 & -0.73 & 7.35 \\ '' & 12010 & 5.8440 & 1.746 & -0.37 &
  -0.03 & -0.04 & -0.30 & -0.04 &  0.01 & -0.11 & -0.65 & 7.64 \\ '' &
  11540 & 5.9253 & 1.971 & -0.33 & 0.00 & -0.04 & -0.26 & -0.03 & 0.01
  & -0.11 & -0.56 & 7.93 \\ '' & 11040 & 5.9987 & 2.223 & -0.30 & 0.04
  & -0.03 & -0.20 & -0.02 & 0.02  & -0.11 & -0.46 & 8.21 \\ '' & 10520
  & 6.0664  & 2.511 &  -0.29 & 0.07  & -0.01 & -0.14  & 0.00 &  0.03 &
  -0.10 & -0.36 & 8.48 \\ '' & 10000 & 6.1303 & 2.849 & -0.30 & 0.12 &
  0.02 &  -0.06 & 0.03 & 0.04  & -0.10 & -0.26  & 8.77 \\ ''  & 9500 &
  6.1927 & 3.279 & -0.33 & 0.16 &  0.05 & 0.06 & 0.06 & 0.06 & -0.10 &
  -0.20 & 9.08 \\  '' & 9050 & 6.2590 & 3.974 & -0.36  & 0.19 & 0.09 &
  0.18 & 0.09  & 0.07 & -0.09 & -0.16  & 9.42 \\ '' &  8660 & 6.3336 &
  5.723 & -0.39 &  0.22 & 0.12 & 0.30 & 0.12 & 0.09  & -0.08 & -0.14 &
  9.78 \\ ''  & 8230 & 6.3987 & 8.461  & -0.42 & 0.26 &  0.15 & 0.44 &
  0.15 & 0.12 & -0.07 & -0.12 & 10.14 \\'' & 7330  & 6.5018 & 15.655 & 
 -0.45  & 0.34 & 0.23 &  0.77 & 0.23 &
  0.17 & -0.04 & -0.10 & 10.88 \\  '' & 6880 & 6.5439 & 20.326 & -0.42
  & 0.40  & 0.26 & 0.95 &  0.27 & 0.20 &  -0.02 & -0.09 &  11.26 \\ \\
  0.230 &  30010 & 6.4188 &  0.203 & -1.06 &  -0.27 & -0.13  & -0.94 &
  -0.17 & -0.10 & -0.17 & -2.98 &  7.12 \\ '' & 28220 & 6.4678 & 0.203
  & -1.02 &  -0.25 & -0.13 & -0.90  & -0.16 & -0.10 & -0.17  & -2.84 &
  7.37 \\ '' & 26430 & 6.5134 &  0.203 & -0.99 & -0.23 & -0.12 & -0.86
  & -0.15  & -0.09 &  -0.16 & -2.68  & 7.61 \\ ''  & 25000 &  6.5480 &
  0.203 &  -0.96 & -0.22  & -0.12 &  -0.82 & -0.15  & -0.08 &  -0.15 &
  -2.54 & 7.80 \\ '' & 23350 &  6.5868 & 0.203 & -0.92 & -0.20 & -0.11
  & -0.78  & -0.14  & -0.07 &  -0.15 & -2.36  & 8.02  \\ '' &  22030 &
  6.6168 &  0.204 & -0.89 &  -0.19 & -0.10 &  -0.74 & -0.13  & -0.07 &
  -0.15 & -2.21 & 8.19 \\ '' &  20710 & 6.6461 & 0.205 & -0.85 & -0.17
  & -0.10 & -0.70 & -0.12 & -0.06 & -0.14 & -2.05 & 8.37 \\ '' & 19200
  & 6.6727 & 0.214  & -0.79 & -0.15 & -0.09 & -0.65  & -0.11 & -0.05 &
  -0.14 & -1.85 & 8.56 \\ '' &  17860 & 6.6782 & 0.251 & -0.74 & -0.13
  & -0.09 & -0.61 & -0.10 & -0.04 & -0.13 & -1.66 & 8.70 \\ '' & 16620
  & 6.6907 & 0.299  & -0.69 & -0.10 & -0.08 & -0.56  & -0.09 & -0.04 &
  -0.13 & -1.47 & 8.85 \\ '' &  15380 & 6.7116 & 0.345 & -0.62 & -0.08
  & -0.07 & -0.50 & -0.08 & -0.03 & -0.13 & -1.26 & 9.04 \\ '' & 14310
  & 6.7331 & 0.390  & -0.56 & -0.05 & -0.07 & -0.45  & -0.07 & -0.02 &
  -0.12 & -1.07 & 9.21 \\ '' &  13320 & 6.7549 & 0.442 & -0.50 & -0.02
  & -0.06 & -0.39 & -0.06 & -0.01 & -0.12 & -0.89 & 9.40 \\ '' & 12400
  & 6.7768  & 0.501 & -0.44  & 0.02 & -0.05  & -0.33 & -0.04  & 0.01 &
  -0.12 & -0.71 & 9.59 \\ '' & 11530 & 6.7990 & 0.570 & -0.40 & 0.06 &
  -0.03 & -0.25 & -0.02 & 0.02 &  -0.11 & -0.55 & 9.79 \\ '' & 10730 &
  6.8210 & 0.648 & -0.40 & 0.11 & 0.00 & -0.15 & 0.01 & 0.03 & -0.11 &
  -0.38 & 9.99 \\  '' & 9910 & 6.8455 & 0.751 & -0.41  & 0.15 & 0.04 &
  0.00 & 0.05 &  0.05 & -0.10 & -0.26 & 10.28 \\ ''  & 9220 & 6.8699 &
  0.868 & -0.44 &  0.20 & 0.09 & 0.17 & 0.09 & 0.07  & -0.09 & -0.20 &
  10.59 \\ '' &  8600 & 6.8939 & 1.015 & -0.47 & 0.24  & 0.13 & 0.35 &
  0.14 & 0.10 & -0.08 & -0.16 &  10.91 \\ '' & 8070 & 6.9151 & 1.193 &
  -0.48 & 0.28 & 0.17 & 0.51 &  0.17 & 0.13 & -0.06 & -0.13 & 11.21 \\'' 
  & 7000 & 6.9588 & 1.824 & -0.46 &
  0.38 &  0.25 & 0.90 & 0.26  & 0.20 & -0.03  & -0.10 & 11.91  \\ '' &
  6460 & 6.9796 &  2.341 & -0.40 & 0.45 & 0.30 & 1.13  & 0.30 & 0.24 &
  -0.01 & -0.10 & 12.31 \\ '' & 5970 & 6.9984 & 2.930 & -0.31 & 0.54 &
  0.36 &  1.37 & 0.36  & 0.27 & 0.02  & -0.12 &  12.71 \\'' & 5150 &  
  7.0477 & 4.075 & -0.09 & 0.74 & 0.47 &
  1.86 & 0.47  & 0.32 & 0.08 & -0.23  & 13.59 \\ '' &  4810 & 7.0820 &
  4.780 &  0.02 & 0.83 & 0.54  & 2.11 & 0.53  & 0.35 & 0.10  & -0.32 &
  14.07 \\ ''  & 4460 & 7.1098 & 5.792  & 0.14 & 0.93 &  0.60 & 2.29 &
  0.60 & 0.35 &  0.06 & -0.43 & 14.58 \\ '' & 4110  & 7.1265 & 6.876 &
  0.24 & 1.02 & 0.66 & 2.23 & 0.66 & 0.18 & 0.00 & -0.49 & 15.04 \\ ''
  & 3780 & 7.1369 & 7.948 & 0.32 & 1.09 & 0.70 & 1.85 & 0.69 & -0.08 &
  -0.11 & -0.46 & 15.39 \\ '' &  3450 & 7.1443 & 9.157 & 0.39 & 1.15 &
  0.73 & 1.25  & 0.69 & -0.27 & -0.26  & -0.35 & 15.69 \\  '' & 3210 &
  7.1486 & 10.232 & 0.46 & 1.20 & 0.74 & 0.74 & 0.64 & -0.36 & -0.35 &
  -0.22 & 15.89 \\ '' & 2980 &  7.1522 & 11.469 & 0.52 & 1.25 & 0.72 &
  0.27 & 0.51 & -0.38 & -0.43 &  -0.07 & 16.07 \\ '' & 2760 & 7.1552 &
  12.897 & 0.59 & 1.29 & 0.69 &  -0.16 & 0.33 & -0.36 & -0.54 & 0.08 &
  16.26 \\ '' & 2530 & 7.1593 &  15.744 & 0.68 & 1.35 & 0.62 & -0.70 &
  0.04 & -0.29 & -0.69 & 0.26 & 16.46 \\ '' & 2310 & 7.1624 & 18.216 &
  0.78 & 1.40 & 0.55 & -1.27 & -0.34 & -0.20 & -0.87 & 0.45 & 16.69 \\
  \\ 0.336 & 21420 & 7.1810 & 0.045  & -0.88 & -0.16 & -0.10 & -0.73 &
  -0.12 & -0.06 & -0.14 & -2.13 &  9.22 \\ '' & 19620 & 7.1907 & 0.049
  & -0.82 &  -0.13 & -0.10 & -0.67  & -0.11 & -0.05 & -0.14  & -1.90 &
  9.40 \\ '' & 18260 & 7.1978 &  0.065 & -0.78 & -0.11 & -0.09 & -0.62
  & -0.10  & -0.05 &  -0.14 & -1.71  & 9.54 \\ ''  & 17010 &  7.2109 &
  0.090 &  -0.72 & -0.09  & -0.08 &  -0.58 & -0.09  & -0.04 &  -0.13 &
  -1.52 & 9.69 \\ '' & 15840 &  7.2270 & 0.121 & -0.67 & -0.06 & -0.08
  & -0.53  & -0.08  & -0.03 &  -0.13 & -1.33  & 9.85  \\ '' &  14640 &
  7.2464 &  0.161 & -0.61 &  -0.03 & -0.07 &  -0.47 & -0.07  & -0.02 &
  -0.12 & -1.12 & 10.03 \\ '' &  13540 & 7.2658 & 0.210 & -0.55 & 0.01
  & -0.06  & -0.41 &  -0.06 & -0.01  & -0.12 & -0.92  & 10.22 \\  '' &
  12620 & 7.2830 & 0.268 & -0.50 & 0.05 & -0.05 & -0.34 & -0.04 & 0.01
  & -0.12  & -0.75 & 10.40  \\ '' & 11810  & 7.2986 & 0.337  & -0.48 &
  0.08 & -0.03 & -0.27 & -0.02 &  0.01 & -0.12 & -0.59 & 10.56 \\ '' &
  10880 & 7.3163 & 0.445 & -0.48 & 0.12 & 0.00 & -0.16 & 0.01 & 0.03 &
  -0.11 & -0.41 & 10.78 \\ '' &  10020 & 7.3328 & 0.582 & -0.48 & 0.16
  & 0.05 & 0.00 &  0.06 & 0.05 & -0.10 & -0.29 & 11.06  \\ '' & 9370 &
  7.3454 & 0.719 & -0.50 & 0.20 &  0.09 & 0.16 & 0.09 & 0.07 & -0.09 &
  -0.23 & 11.33 \\ '' & 8720 &  7.3573 & 0.889 & -0.52 & 0.23 & 0.13 &
  0.33 & 0.13 &  0.10 & -0.08 & -0.18 & 11.62 \\ ''  & 8120 & 7.3681 &
  1.092 & -0.52 &  0.27 & 0.17 & 0.51 & 0.17 & 0.13  & -0.06 & -0.15 &
  11.93 \\ '' &  7460 & 7.3795 & 1.370 & -0.51 & 0.33  & 0.22 & 0.73 &
  0.22 & 0.17 & -0.04 & -0.12 &  12.29 \\ '' & 6950 & 7.3883 & 1.640 &
  -0.47 & 0.39 & 0.26 & 0.93 &  0.26 & 0.20 & -0.03 & -0.11 & 12.61 \\
  '' &  6390 & 7.3987 & 2.023  & -0.40 & 0.47  & 0.31 & 1.17  & 0.31 &
  0.24 & 0.00 & -0.11 & 13.00 \\  '' & 5870 & 7.4099 & 2.492 & -0.29 &
  0.56 & 0.37 & 1.43 & 0.37 & 0.28 & 0.03 & -0.13 & 13.41 \\'' & 5000 &  
  7.4416 & 3.926 & -0.05 & 0.77 & 0.50 &
  1.96 & 0.50  & 0.34 & 0.08 & -0.26  & 14.32 \\ '' &  4610 & 7.4569 &
  5.361 &  0.07 & 0.88 & 0.57  & 2.18 & 0.57  & 0.35 & 0.06  & -0.38 &
  14.83 \\ ''  & 4240 & 7.4669 & 6.985  & 0.18 & 0.97 &  0.63 & 2.16 &
  0.63 & 0.20 & -0.01 & -0.44 &  15.29 \\ '' & 3890 & 7.4727 & 8.597 &
  0.26 & 1.05 & 0.68 & 1.80 &  0.67 & -0.08 & -0.10 & -0.42 & 15.65 \\
  '' &  3610 & 7.4760  & 9.997 & 0.32  & 1.10 &  0.70 & 1.28 &  0.68 &
  -0.26 & -0.25 & -0.34 & 15.90 \\  '' & 3300 & 7.4790 & 11.832 & 0.40
  & 1.16 & 0.72 & 0.76 & 0.64 &  -0.36 & -0.34 & -0.21 & 16.17 \\ '' &
  3020 & 7.4813 & 14.033 & 0.47 &  1.21 & 0.70 & 0.22 & 0.51 & -0.39 &
  -0.43 & -0.05 & 16.39 \\ 
\hline
\end{tabular}
\label{tab.z2}

{\small Ages are counted from the end of mass transfer.}
\end{minipage}
\end{scriptsize}
\end{table*}

\begin{table*}
\centering
\begin{scriptsize}
\begin{minipage}{135mm}
\setcounter{table}{2}
\caption{Continued.}
\begin{tabular}{@{}ccccccccccccc@{}}
\hline $M_*/{\rm M_{\odot}}$  & \teff & $Log (g)$ &  $Age$ (Gyr) & U-B
&B-V & V-R & V-K & R-I & J-H & H-K & BC & M$_V$ \\ \hline

 \\ 0.449 & 36370 & 7.2604 & 0.002 & -1.16 & -0.28 & -0.14 &
  -1.00 & -0.18 & -0.12 & -0.17 &  -3.45 & 8.13 \\ '' & 34060 & 7.2999
  & 0.004 &  -1.13 & -0.27 & -0.14  & -0.99 & -0.18 & -0.11  & -0.17 &
  -3.27 & 8.34 \\ '' & 32270 &  7.3577 & 0.011 & -1.11 & -0.26 & -0.14
  & -0.97  & -0.17  & -0.11 &  -0.17 & -3.13  & 8.58  \\ '' &  30310 &
  7.4037 &  0.023 & -1.09 &  -0.25 & -0.13 &  -0.94 & -0.17  & -0.10 &
  -0.17 & -2.98 & 8.82 \\ '' &  28220 & 7.4342 & 0.035 & -1.05 & -0.23
  & -0.13 & -0.90 & -0.16 & -0.10 & -0.17 & -2.82 & 9.04 \\ '' & 26220
  & 7.4605 & 0.049  & -1.01 & -0.21 & -0.12 & -0.86  & -0.15 & -0.09 &
  -0.16 & -2.64 & 9.24 \\ '' &  24310 & 7.4844 & 0.068 & -0.97 & -0.19
  & -0.12 & -0.81 & -0.14 & -0.08 & -0.15 & -2.45 & 9.44 \\ '' & 22530
  & 7.5065 & 0.093  & -0.93 & -0.16 & -0.11 & -0.76  & -0.13 & -0.07 &
  -0.15 & -2.25 & 9.63 \\ '' &  20860 & 7.5270 & 0.128 & -0.88 & -0.14
  & -0.10 & -0.71 & -0.12 & -0.06 & -0.14 & -2.05 & 9.81 \\ '' & 19290
  & 7.5462 & 0.174  & -0.83 & -0.11 & -0.10 & -0.66  & -0.11 & -0.05 &
  -0.14 & -1.84 & 9.99 \\ '' &  17830 & 7.5638 & 0.237 & -0.78 & -0.09
  & -0.09  & -0.61 &  -0.10 & -0.04  & -0.14 & -1.63  & 10.17 \\  '' &
  16470 &  7.5796 & 0.320 &  -0.72 & -0.06 &  -0.08 & -0.56  & -0.09 &
  -0.03 & -0.13 & -1.42 & 10.34 \\ '' & 15200 & 7.5938 & 0.429 & -0.66
  & -0.02 & -0.07  & -0.50 & -0.07 & -0.02 & -0.13  & -1.21 & 10.51 \\
  '' & 14030 & 7.6063 & 0.570 & -0.60 & 0.01 & -0.07 & -0.44 & -0.06 &
  -0.01 & -0.12 & -1.00 & 10.69 \\ '' & 12920 & 7.6173 & 0.749 & -0.55
  & 0.06 & -0.05 & -0.36 & -0.04  & 0.00 & -0.12 & -0.80 & 10.87 \\ ''
  & 11890 &  7.6273 & 0.982 & -0.54  & 0.09 & -0.03 & -0.28  & -0.02 &
  0.01 & -0.12 & -0.59 & 11.04 \\  '' & 10910 & 7.6365 & 1.289 & -0.53
  & 0.13 & 0.01 & -0.15 & 0.02 &  0.03 & -0.11 & -0.41 & 11.26 \\ '' &
  10030 & 7.6452 & 1.672 & -0.53 &  0.17 & 0.05 & 0.02 & 0.06 & 0.05 &
  -0.10 & -0.30 & 11.54 \\ '' & 9220 & 7.6532 & 2.148 & -0.54 & 0.21 &
  0.10 &  0.21 & 0.11 & 0.08  & -0.09 & -0.23  & 11.85 \\ ''  & 8460 &
  7.6606 & 2.728 & -0.55 & 0.25 &  0.15 & 0.41 & 0.15 & 0.11 & -0.07 &
  -0.18 & 12.19 \\ '' & 7770 &  7.6674 & 3.349 & -0.54 & 0.30 & 0.20 &
  0.63 & 0.20 &  0.15 & -0.05 & -0.14 & 12.54 \\ ''  & 7130 & 7.6741 &
  3.969 & -0.49 &  0.36 & 0.24 & 0.86 & 0.25 & 0.19  & -0.03 & -0.12 &
  12.91 \\ '' &  6540 & 7.6806 & 4.611 & -0.42 & 0.44  & 0.30 & 1.10 &
  0.30 & 0.23 & -0.01 & -0.10 &  13.28 \\ '' & 6020 & 7.6874 & 5.330 &
  -0.32 & 0.54 &  0.35 & 1.34 & 0.35 & 0.26 & 0.02  & -0.12 & 13.67 \\
  '' &  5540 & 7.6959 & 6.189  & -0.21 & 0.63  & 0.41 & 1.59  & 0.41 &
  0.29 & 0.05 & -0.14 & 14.07 \\  '' & 5110 & 7.7067 & 7.472 & -0.09 &
  0.74 & 0.48 & 1.88 & 0.48 & 0.33 & 0.07 & -0.24 & 14.55 \\ '' & 4710
  & 7.7160 & 9.455 & 0.04 & 0.85 &  0.55 & 2.10 & 0.55 & 0.34 & 0.05 &
  -0.34 & 15.04 \\ '' & 4330 &  7.7221 & 11.582 & 0.14 & 0.94 & 0.61 &
  2.10 & 0.61 &  0.20 & -0.01 & -0.41 & 15.49 \\ ''  & 3970 & 7.7258 &
  13.651 & 0.22 & 1.02 & 0.66 &  1.79 & 0.65 & -0.06 & -0.09 & -0.40 &
  15.85 \\ '' &  3640 & 7.7282 & 15.838 & 0.30 & 1.08  & 0.69 & 1.19 &
  0.66 & -0.28 & -0.26 & -0.31 & 16.16 \\
\hline
\end{tabular}
\label{tab.z2}

{\small Ages are counted from the end of mass transfer.} 

\end{minipage}
\end{scriptsize}
\end{table*}

\bsp

\label{lastpage}

\end{document}